\documentclass[aps,pre,twocolumn,amsmath,amssymb,superscriptaddress,reprint,longbibliography]{revtex4-1}
\usepackage{amsfonts,amsmath,amssymb,bm}
\usepackage{graphicx,graphics,float}
\usepackage{bm}
\usepackage{amssymb}
\usepackage{colordvi}
\usepackage{graphicx}
\usepackage{color}
\usepackage[colorlinks=true,linkcolor=blue,citecolor=blue,urlcolor=blue]{hyperref}%
\usepackage{hyperref}
\usepackage{comment}
\usepackage{harpoon}

\begin{document}

\title{Optimal efficiency and power and their trade-off in three-terminal quantum thermoelectric engines with two output electric currents}

\author{Jincheng Lu}%\email{These authors contributed equally.}%\email{jincheng_lu@qq.com}
\affiliation{School of physical science and technology \&
Collaborative Innovation Center of Suzhou Nano Science and Technology, Soochow University, Suzhou 215006, China.}

\author{Yefeng Liu}%\email{These authors contributed equally.}
\affiliation{School of physical science and technology \&
Collaborative Innovation Center of Suzhou Nano Science and Technology, Soochow University, Suzhou 215006, China.}

\author{Rongqian Wang}%\email{rongqianwang@sina.com}
\affiliation{School of physical science and technology \&
Collaborative Innovation Center of Suzhou Nano Science and Technology, Soochow University, Suzhou 215006, China.}

\author{Chen Wang}%\email{wangchenyifang@gmail.com}
\address{Department of Physics, Zhejiang Normal University, Jinhua, Zhejiang 321004, China}

\author{Jian-Hua Jiang}\email{jianhuajiang@suda.edu.cn}
\affiliation{School of physical science and technology \&
Collaborative Innovation Center of Suzhou Nano Science and Technology, Soochow University, Suzhou 215006, China.}

\date{\today}
\begin{abstract}
We establish a theory of optimal efficiency and power for three-terminal thermoelectric engines which have
two independent output electric currents and one input heat current. This set-up goes beyond the conventional heat
engines with only one output electric current. For such a set-up, we derive the optimal
efficiency and power and their trade-off for three-terminal heat engines with and without time-reversal
symmetry. The formalism goes beyond the known optimal efficiency and power for systems with or without time-reversal
symmetry, showing interesting features that have not been revealed before. A concrete example of quantum-dot heat engine
is studied to show that the current set-up can have much improved efficiency and power compared with previous
set-ups with only one output electric current. Our analytical results also apply for
thermoelectric heat engines with multiple output electric currents, providing an alternative scheme toward future
high-performance thermoelectric materials.
\end{abstract}

\pacs{05.70.Ln, 84.60.-h, 88.05.De, 88.05.Bc}

\maketitle

\section{Introduction}
Thermoelectric phenomena have attracted lots of research attention because of their relevance to 
fundamental physics and the state-of-art energy applications~\cite{butcher1990,DubiRMP,RenRMP,Nanotechnology,JiangCRP,BENENTI20171}.
The understanding of fundamental thermodynamic constraints on the efficiency and power of nanoscale thermoelectric 
devices is a subject of wide-spread interest in the past decades~\cite{JiangOra,JiangPRE,PhysRevE.93.042112,Proesmans,Constancy,WhitneyPRL,WhitneyPRB,verley2014unlikely,Polettini,JiangPRL,Brownian-Duet}.
Recent theoretical~\cite{JiangOra,JiangPRE,PhysRevE.93.042112,Proesmans,Constancy,WhitneyPRL,WhitneyPRB,verley2014unlikely,Polettini,JiangPRL,Brownian-Duet,David2011PRB,Rafael,Jiang2012,Sothmann-Re,Jiang2013,Simine,Sothmann-QW,JiangNJP,Mazza-separation,Udo-MT,BijayJiang,JiangSR,Yamamoto,JiangBijayPRB17,Brandner2018,JiangNearfield,Polettini,Naoto,Brandner2018,MyPRB}
and experimental~\cite{petersson,hwang,Exper,Thier2015,roche2015,cui2018peltier} studies on thermoelectric
phenomena in mesoscopic systems have renewed the fundamental understanding on
thermoelectric transport and energy conversion. Several concepts, such as reversal thermoelectric energy 
conversion~\cite{PhysRevLett.89.116801,PhysRevLett.94.096601}, inelastic thermoelectric transport~\cite{David2011PRB,Rafael,Jiang2012,Sothmann-Re,Jiang2013,Simine,Sothmann-QW,JiangNJP,JiangCRP}, fundamental bounds on the optimal efficiency and power~\cite{JiangOra,JiangPRE,PhysRevE.93.042112,Proesmans,Constancy,WhitneyPRL,WhitneyPRB,Iyynonlin}, 
universal fluctuations of energy efficiency~\cite{verley2014unlikely,Polettini,JiangPRL,Brownian-Duet}, cooperative effects~\cite{JiangPRE,JiangJAP,MyJAP}, and nonlinear effects~\cite{Jiang2017} were proposed. In particular,
with the seminal works by Benenti~{\sl et al.}~\cite{Saito2011} and 
later by Brandner~{\sl et al.}~\cite{StrongBounds} mesoscopic thermoelectric heat engines with broken time-reversal symmetry 
have gained much interest, particularly in multi-terminal transport configurations~\cite{saito,Balachandran,Udo-MT,BrandnerPRE,UdoPRL,BENENTI20171} where
thermoelectric engines with asymmetric Onsager transport coefficients are studied in the set-up with one heat current 
input and one electric current output.

In the linear-response regime, the transport property of a thermoelectric engine is described by the following equation,
\begin{align}
  \left( \begin{array}{c}
      I_e\\ I_{Q} \end{array}\right) =
  \left( \begin{array}{cccc}
      G & L_1 \\
      L_2 & K
    \end{array}\right) \left(\begin{array}{c}
       V \\  \frac{T_h-T_c}{T_h} \end{array}\right) .
\end{align}
where $I^e$ and $I^{Q}$ are the charge and heat currents, $G$ and $K$ are the charge and heat conductivity, 
respectively. $L_1$ and $L_2$ describe the Seebeck and Peltier effects, respectively, $V$ is the voltage bias
across the device, $T_h$ and $T_c$ are the temperatures of the hot and cold reservoirs, respectively. In 
time-reversal broken multi-terminal systems the two coefficients $L_1$ and $L_2$ can be different~\cite{Saito2011}, 
though they are often identical for time-reversal symmetric thermoelectric devices. Thermoelectric efficiency is defined
as $\eta=-I_eV/I_Q$ with $I_eV<0$ (power output) and $I_Q>0$ (heat consumption). As shown in Ref.~\cite{Saito2011}, 
for a thermoelectric heat engine described by the above equation, the maximal efficiency and efficiency at maximal power 
of the thermoelectric heat engine are given by
\begin{align}
\eta_{\max}=\eta_C r_{12} \frac{\sqrt{ZT+1}-1}{\sqrt{ZT+1}+1}, \quad \eta(W_{\max})=\frac{\eta_C}{2}\frac{r_{12} ZT}{2+ZT}
\label{casati}
\end{align}
respectively, where $\eta_C= \frac{T_h-T_c}{T_h}$ is the Carnot efficiency and
\begin{align}
ZT = \frac{L_1L_2}{GK-L_1L_2}, \quad r_{12}= \frac{L_1}{L_2} ,
\end{align}
are the thermoelectric figure-of-merit and the partition ratio between the two off-diagonal elements.
For a time-reversal symmetric macroscopic system (length $l$ and cross-section area ${\cal A}$), the above equations 
comes back to the more familiar form, $r_{12}=1$ and the figure-of-merit $ZT=\frac{\sigma S^2 T}{\kappa}$ where 
$\sigma=Gl/{\cal A}$ is the conductivity, $S=L/(TG)$ is the Seebeck coefficient, $\kappa = (K-L_1L_2/G)l/({\cal A}T)$ is 
the thermal conductivity. Eq.~(\ref{casati}) also gives guidance to exceed the so-called Curzon-Ahlborn
limit~\cite{Curzon-limit} $\eta_{CA}$ for the efficiency at maximal power (in the linear-response regime 
$\eta_{CA}=\frac{\eta_C}{2}$).

However, the existing studies on thermoelectric energy conversion in time-reversal broken systems are
restricted to the situation with only one output electric current~\cite{UdoPRL,BrandnerPRE}. 
Even in multi-terminal systems, other electric currents are suppressed by tuning the electrochemical 
potentials and temperatures~\cite{Udo-MT}. Such artificial constraints limit the study of
thermoelectric energy conversion in generic multi-terminal mesoscopic systems. In this work, we go beyond such
constraints by studying multi-terminal mesoscopic systems connected with two heat baths while there can be
multiple output electric currents using multiple electrodes. For concreteness, we study a three-terminal 
thermoelectric heat engine with two output electric currents. We find that by going beyond previous limitation 
of only one output electric current, the efficiency and power can be significantly improved. We derive the 
analytical expressions for the optimal efficiency and power for the set-up with multiple output electric currents 
and find their trade-off relations~\cite{JiangPRE,BrandnerPRE,Proesmans,Constancy,WhitneyPRL} in
the linear-response regime. Our study shows that multi-terminal mesoscopic systems have the potential of
achieving higher energy efficiency and larger output power than two-terminal systems, particularly in the set-ups with
multiple output electric currents.

The main part of the paper is organized as follows. In Sec.~\ref{sec:3terminals}, we introduce the mesoscopic transport
model. In Sec.~\ref{se:eta_m}, we obtain the optimal efficiency and power, and derive the relations between the
maximum efficiency, maximum power, efficiency at maximal power and power at maximal efficiency in the
linear-response regime. In Sec.~\ref{upper-bound}, we deduce the bounds for the optimal efficiency and power
in the linear-response regime. In Sec.~\ref{non-system}, we analyze the efficiency and power of a triple-quantum-dot
three-terminal mesoscopic system. We conclude and remark for future studies in Sec.~\ref{conclusion}.

\section{Theoretical model and formulation}\label{sec:3terminals}

\begin{figure}
\begin{center}
\centering \includegraphics[width=6.6cm]{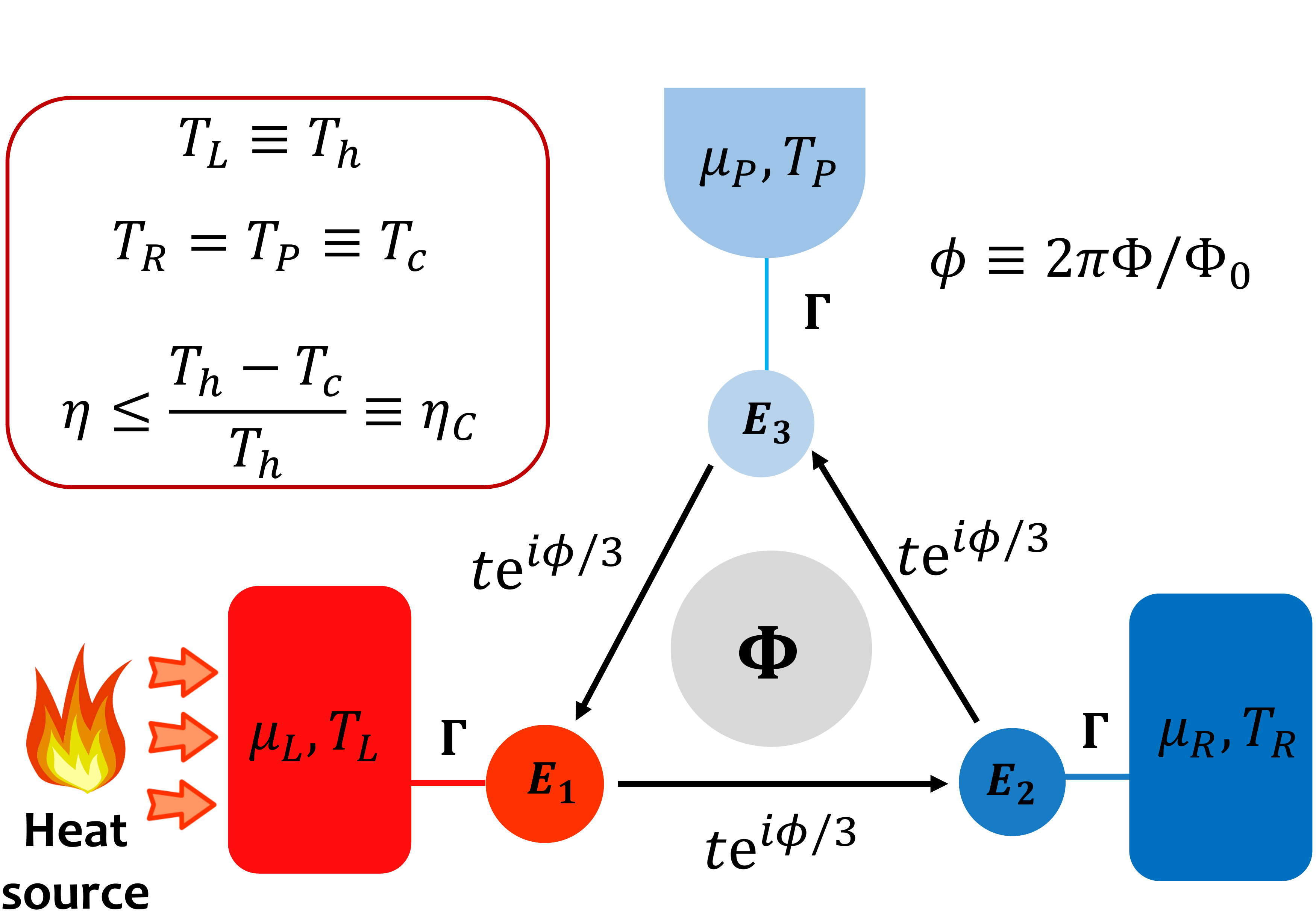}
\caption{(Color online) Schematic of a triple-quantum-dot thermoelectric heat engine with a magnetic flux $\Phi$. Three 
quantum dots (energy levels are denoted as $E_i$, $i=L,R,P$) are connected to three electrodes. The electrochemical 
potentials and temperatures of the electrodes labeled in the figure. The tunneling rate between the dots and the electrodes
is denoted by $\Gamma$. We consider the situation when both the $R$ and $P$ electrodes are in contact with a cold
heat bath with temperature $T_c$, while the $L$ electrode is connected to a hot heat bath with temperature $T_h$.
}
\label{fig1}
\end{center}
\end{figure}

As shown in Fig.~\ref{fig1}, we consider a nanoscale thermoelectric device consisting of three quantum dots (QDs)
coupled to three electrodes. This is a minimal model to demonstrate the set-up with two output electric currents.
Although this model has been studied before~\cite{saito,Balachandran,Udo-MT,UdoPRL}, the configuration with
two output electric currents has never been studied in the time-reversal symmetry broken regime. This model is
valid when the Coulomb interaction in the QDs can be neglected~\cite{Buttiker}. Each QD is coupled to the
nearby reservoir and we thus employ the indices $1/2/3$ to label the leads $L/R/P$, respectively~\cite{JiangPRL}.

Hoppings between QDs are affected by the magnetic flux $\Phi$ piercing through the device at the center with
the phase $\phi/3$ assigned to each of the hoppings ($\phi=2\pi\Phi/\Phi_0$ where $\Phi_0$ is flux quantum).
The system is described by the Hamiltonian~\cite{saito}
\begin{align}
\hat{H}=\hat{H}_{qd}+\hat{H}_{lead}+\hat{H}_{tun},
\end{align}
where
\begin{align}
\hat{H}_{qd}=\sum_{i=1,2,3} E_i d_i^\dagger d_i + (t e^{i\phi/3}d_{i+1}^\dagger d_i + {\rm H.c.}),
\end{align}
\begin{align}
\hat{H}_{lead}=\sum_{i=1,2,3}\sum_k\varepsilon_{k} c^\dagger_{ik} c_{ik},
\end{align}
\begin{align}
\hat{H}_{tun}=\sum_{i,k}V_{ik}d^\dagger_ic_{ik}+{\rm H.c.}.
\end{align}
Here, $d_i^{\dag}$ and $d_i$ create and annihilate an electron in the $i$th QD with an energy $E_i$, respectively,
$t$ is the tunneling amplitude between the QDs. $c_{ik}^{\dag}$ and $c_{ik}$ create and annihilate an electron in
the $i$-th electrode with the energy $E_i$ ($i=1,2,3$).

The chemical potential and temperature of three reservoirs are denoted by $\mu_i$ and $T_i$ ($i=L,R,P$), respectively.
For each reservoir, there are an electric and a heat currents flowing out of the reservoir. In total there are six currents.
However, only four of them are independent, due to charge and energy conservation~\cite{saito}. We choose the
charge and heat currents flowing out of the $L$ and $P$ reservoirs as the independent currents which are denoted as
$I_e^i$ and $I_Q^i$ ($i=L,P$), respectively. The corresponding thermodynamic forces are
\begin{equation}
F_e^i=\frac{\mu_i-\mu_R}{e}, \quad F_Q^i=\frac{T_i-T_R}{T_i} \quad  (i=L,P).
\end{equation}
where $e<0$ is the electronic charge. We focus on the set-up where $L$ reservoir is connected to the hot bath
and the $R$ and $P$ reservoirs are connected to the cold bath, i.e., $T_L=T_h$ and $T_P=T_R=T_c$. There are
two independent output electric currents, $I_e^L$ and $I_e^P$ (i.e., the charge currents flowing out of the $L$ and
$P$ reservoirs), whereas there is only one input heat current $I_Q\equiv I_Q^L$ (i.e., the heat current flowing out of the hot
reservoir $L$) with the corresponding force $F_Q\equiv F_Q^L$.

With such a set-up, the phenomenological Onsager transport equation is written in the linear-response regime as
\begin{equation}
\begin{pmatrix}
\vec{I}_{e} \\
{I}_{Q} \\
\end{pmatrix}
=
\begin{pmatrix}
\hat{\mathcal M}_{ee} & \hat{\mathcal M}_{e Q}   \\
\hat{\mathcal M}_{Qe} & {\mathcal M}_{QQ}  \\
\end{pmatrix}
\begin{pmatrix}
\vec{F}_e \\
{F}_Q \\
\end{pmatrix},
\end{equation}
where the symbols $e$ and $Q$ are used to abbreviate the indices of forces and currents for charge and heat, respectively
(i.e., $\vec{I}_e = (I_e^L,I_e^P)^T$, $\vec{F}_e = (F_e^L,F_e^P)^T$, $I_Q\equiv I_Q^L$, and $F_Q\equiv F_Q^L$; here the
superscript $T$ stands for vector/matrix transpose). $\hat{\mathcal M}_{ee}$ denotes the $2\times 2$ charge conductivity
tensor, the $2\times 1$ matrix $\hat{\mathcal M}_{e Q}$ describes the Seebeck effect, while the matrix $\hat{\mathcal M}_{Qe}$
describes the Peltier effect. The $1\times 1$ matrix (scalar) ${\mathcal M}_{QQ}$ represents the heat conductivity.
For systems with time-reversal symmetry (e.g., $\phi=0, \pi$), Onsager's reciprocal relation gives
$\hat{\mathcal M}_{e Q}=\hat{\mathcal M}_{Qe}^T$. In contrast, for time-reversal broken systems, they are
not equal to each other.

The output power and energy efficiency of
the thermoelectric heat engine are written respectively as
\begin{equation}
W = -\vec{I}_e^T\vec{F}_e= -(\vec{F}_e^T \hat{\mathcal M}_{ee} \vec{F}_e + \vec{F}_e^T \hat{\mathcal M}_{e Q} {F}_Q)>0 ,
\end{equation}
and
\begin{equation}
\eta = \frac{W}{{I}_{Q}}=-\frac{\vec{F}_e^T \hat{\mathcal M}_{ee} \vec{F}_e + \vec{F}_e^T \hat{\mathcal M}_{e Q} {F}_Q}{\hat{\mathcal M}_{Qe} \vec{F}_e + {\mathcal M}_{QQ}{F}_Q}\le\eta_C .
\label{eq:eta}
\end{equation}
Here $\eta_C=1-T_c/T_h= {\mathcal {F}}_Q$ is the Carnot efficiency which is the absolute upper bound for the
attainable energy efficiency due to the second-law of thermodynamics of thermodynamics.

\section{Maximal efficiency and power for time-reversal broken systems}\label{se:eta_m}
We note that in the linear-response regime the energy efficiency is invariant under the scaling transformation
$\vec{F}_e\to a \vec{F}_e$ and $F_Q\to a F_Q$ with $a$ being an arbitrary constant. In comparison,
the output power scales as $W\to a^2 W$. We can then fix $F_Q$ and obtain the maximal energy efficiency
by solving the following differential equation,
\begin{equation}
\frac{\partial\eta}{\partial \vec{F}_e} = 0 .
\end{equation}
We obtain that
\begin{equation}
\vec{F}_e = -\frac{1}{2}\left[\eta_{\max}\left(\overline{\hat{\mathcal M}_{ee}}\right)^{-1}\hat{\mathcal M}^T_{Qe}{F}_Q + \left(\overline{\hat{\mathcal M}_{ee}}\right)^{-1}\hat{\mathcal M}_{eQ}{F}_Q \right].
\label{eq:FQ}
\end{equation}
Here we define
\begin{equation}
\overline{\hat{\mathcal M}_{ee}}\equiv \frac{1}{2}(\hat{\mathcal M}_{ee}+\hat{\mathcal M}^T_{ee})
\end{equation}
as the symmetric charge conductivity tensor. Inserting Eq.~\eqref{eq:FQ} into Eq.~\eqref{eq:eta}, we arrive at
\begin{equation}
\eta_{\max} = \eta_C \frac{\lambda_1-\lambda_2(\eta_{\max}/\eta_C)^2}{4-2(\lambda_2(\eta_{\max}/\eta_C)+\lambda_3)} .
\end{equation}
Solving the above quadratic equation, we obtain the maximal efficiency as
\begin{equation}
\eta_{\max}=\eta_C\frac{2-\lambda_3-\sqrt{(2-\lambda_3)^2-\lambda_1\lambda_2}}{\lambda_2}.
~\label{eq:eta_max}
\end{equation}
Here,
\begin{subequations}
\begin{align}
&\lambda_1\equiv \hat{\mathcal M}^T_{eQ}\left(\overline{\hat{\mathcal M}_{ee}}\right)^{-1}\hat{\mathcal M}_{eQ}{\mathcal M}^{-1}_{QQ},\\
&\lambda_2\equiv \hat{\mathcal M}_{Qe}\left(\overline{\hat{\mathcal M}_{ee}}\right)^{-1}\hat{\mathcal M}^T_{Qe}{\mathcal M}^{-1}_{QQ},\\
&\lambda_3\equiv \hat{\mathcal M}_{Qe}\left(\overline{\hat{\mathcal M}_{ee}}\right)^{-1}\hat{\mathcal M}_{eQ}{\mathcal M}^{-1}_{QQ}.
\end{align}\label{eq:lambda}
\end{subequations}
are three dimensionless parameters that characterize the thermoelectric transport properties of the system.
The output power at maximum efficiency is
\begin{equation}
W(\eta_{\max})=
W_0\left[\lambda_1-\lambda_2\left(\frac{\eta_{\max}}{\eta_C}\right)^2\right], \quad W_0\equiv \frac{1}{4}{\mathcal M}_{QQ}{F}^2_Q.
~\label{eq:W_etamax}
\end{equation}

Similarly, we can obtain the maximal output power with fixed $F_Q$ by solving the following equation
\begin{equation}
\frac{\partial W}{\partial \vec{F}_e} =0 ,
\end{equation}
which yields
\begin{equation}
W_{\max}=\lambda_1W_0 .
~\label{eq:W_max}
\end{equation}
Meanwhile, the efficiency at maximum output power is~\cite{Van2005,AI-PRL2012}
\begin{equation}
\eta(W_{\max})=\eta_C\frac{\lambda_1}{4-2\lambda_3}.
~\label{eq:eta_Wmax}
\end{equation}

Comparing the energy efficiency and output power for the above two optimization schemes, we find that
\begin{equation}
\frac{\eta_{\max}}{\eta(W_{\max})} = 1+\frac{\lambda_2}{\lambda_1}\left(\frac{\eta_{\max}}{\eta_C}\right)^2,
~\label{eq:eta_Wmax}
\end{equation}
and
\begin{equation}
\frac{W(\eta_{\max})}{W_{\max}}=1-\frac{\lambda_2}{\lambda_1}\left(\frac{\eta_{\max}}{\eta_C}\right)^2.
~\label{eq:W_etamax}
\end{equation}
The above trade-off relations between the optimization of the efficiency and power is presented graphically in Fig.~2.
These relations also reveal two important properties: First, the performance of the thermoelectric engine
is better when $\lambda_2<\lambda_1$ compared with the situation with $\lambda_2>\lambda_1$. In addition,
when $\lambda_2<\lambda_1$, the efficiency at maximal output power can possibly exceed the Curzon-Ahlborn
limit~\cite{Curzon-limit} in the linear-response $\eta_{CA}=\eta_C/2$. Second, for $\lambda_2<\lambda_1$, the
second-law of thermodynamics does not forbid the Carnot efficiency at finite output power. Although there have
been many debates on such a possibility~\cite{Saito2011,PhysRevLett.117.190601,PhysRevLett.116.160601,PhysRevLett.111.050601,PhysRevLett.110.070604,PhysRevLett.121.120601}, our study here opens a  
regime for further investigation of such an issue in quantum heat engines without the limitation of having only 
one electric and one heat currents.

%%%%%%%%%%%%%%%%%%%%%%%%%%%%%%%%%%%%%%%%%%%%%%%%%%%%%%%%%%%%%%%%%%
\begin{figure}[htb]
\begin{center}
\centering\includegraphics[width=4.2cm]{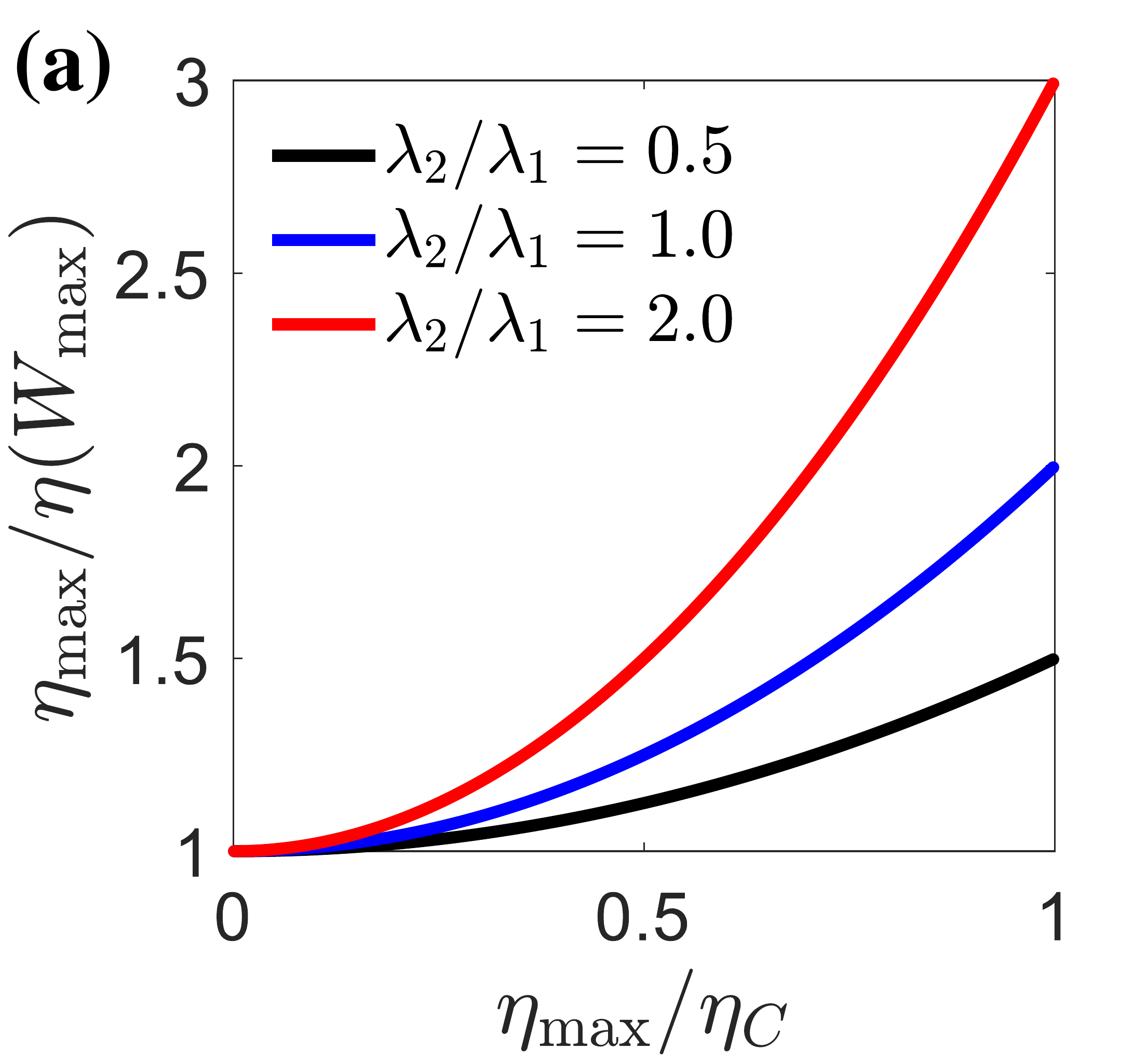}\hspace{0.2cm}\includegraphics[width=4.2cm]{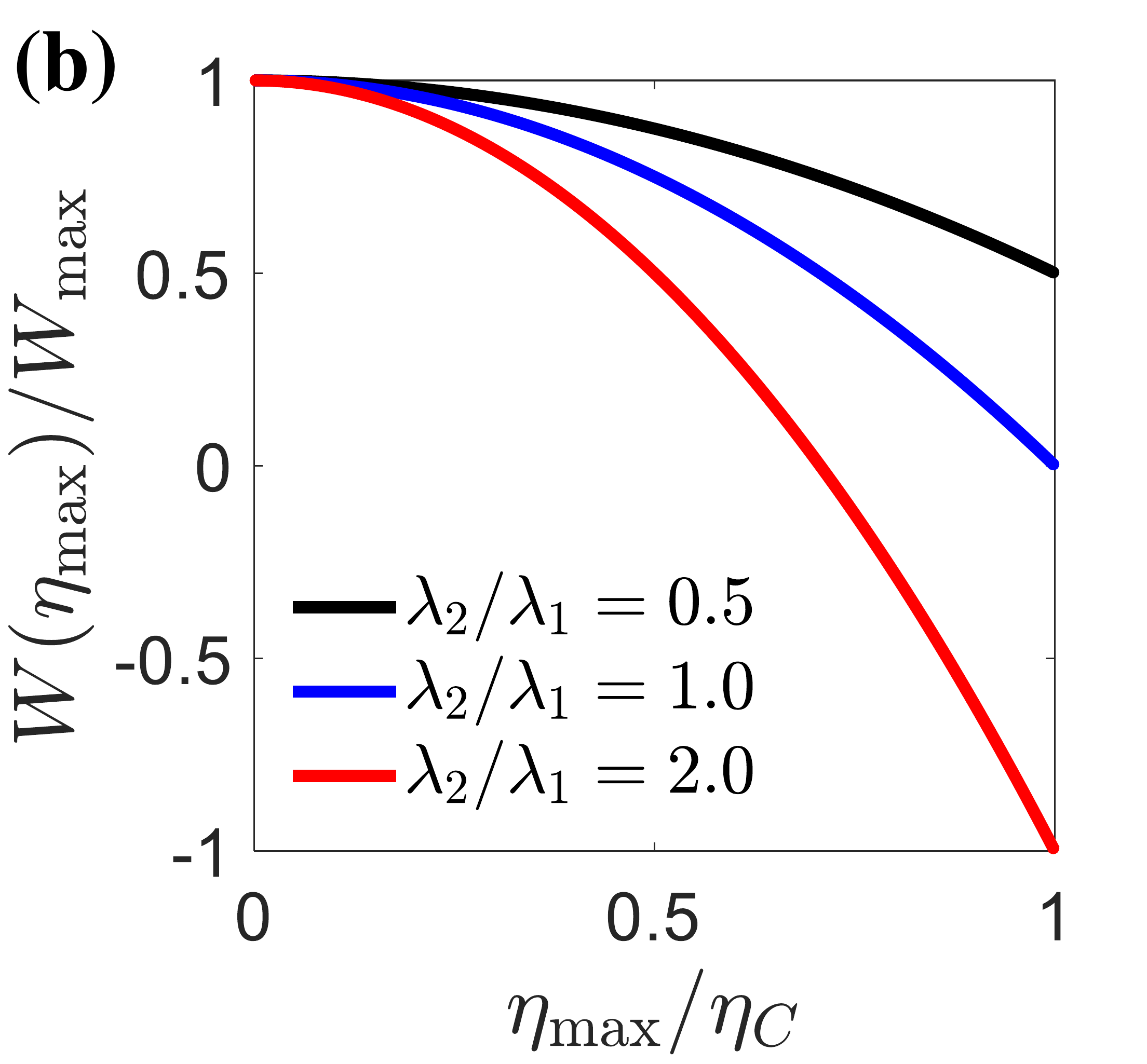}
\caption{(Color online) Trade-off relations for optimal efficiency and power, Eqs.~(\ref{eq:eta_Wmax}) and (\ref{eq:W_etamax}). (a) $\eta_{\max}/\eta(W_{\max})$ and (b) $W(\eta_{\max})/W_{\max}$ as functions of $\eta_{\max}/\eta_C$ for various $\lambda_2/\lambda_1$.}
\label{fig:ratios}
\end{center}
\end{figure}
%%%%%%%%%%%%%%%%%%%%%%%%%%%%%%%%%%%%%%%%%%%%%%%%%%%%%%%%%%%%%%%%%%

We now make two important remarks. First, the above results are valid for the situation with multiple output electric currents.
This can be readily verified through the vectorial (matrix) formulation used in the above discussions.
Second, the second-law of thermodynamics imposes the following constraints on the dimensionless parameters,
\begin{equation}
\lambda_1\ge0, \quad \lambda_2\ge0, \quad \lambda_1+\lambda_2+2\lambda_3\le4.
~\label{eq:Bound1}
\end{equation}
The derivation of the above constraints goes as follows. The entropy production rate associated with
the thermoelectric transport is~\cite{JiangOra}
\begin{equation}
\begin{aligned}
T_R\dot{\mathcal{S}}&=\vec{I}_{e}^T \vec{F}_{e} + {I}_{Q} {F}_Q\\
&=\begin{pmatrix}
\vec{F}_e \,\,\,
{F}_Q
\end{pmatrix}
\begin{pmatrix}
\hat{\mathcal M}_{ee} & \hat{\mathcal M}_{eQ}   \\
\hat{\mathcal M}_{Qe} & {\mathcal M}_{QQ}  \\
\end{pmatrix}
\begin{pmatrix}
\vec{F}_e \\
{F}_Q \\
\end{pmatrix}
\end{aligned}
\end{equation}
The second-law of thermodynamics requires $\dot{\mathcal{S}}\ge 0$ for all values of $\vec{F}_e$ and ${F}_Q$,
which is equivalent to require the following matrix to be positive semi-definite,
\begin{equation}
\begin{aligned}
\begin{pmatrix}
\overline{\hat{\mathcal M}_{ee}} & \frac{\hat{\mathcal M}^T_{eQ}+\hat{\mathcal M}_{Qe}}{2}   \\
\frac{\hat{\mathcal M}^T_{Qe}+\hat{\mathcal M}_{eQ}}{2} & {\mathcal M}_{QQ}  \\
\end{pmatrix} .
\end{aligned}
\end{equation}
Therefore, ${\mathcal M}_{QQ}\ge0$ and the matrix $\overline{\hat{\mathcal M}_{ee}}$ is positive semi-definite. In addition,
the determinant of the above matrix is positive semi-definite which yields
{\small{
\begin{equation}
\begin{aligned}
\left|\overline{\hat{\mathcal M}_{ee}}\right|\left|{\mathcal M}_{QQ}-\frac{\hat{\mathcal M}^T_{eQ}+\hat{\mathcal M}_{Qe}}{2}\left(\overline{\hat{\mathcal M}_{ee}}\right)^{-1} \frac{\hat{\mathcal M}^T_{Qe}+\hat{\mathcal M}_{eQ}}{2} \right| \ge 0 ,
\end{aligned}
\end{equation}}}
where $|\,|$ is the determinant of the matrix. From these positive semi-definite properties, one can deduce
Eq.~(\ref{eq:Bound1}) straightforwardly.

We now compare our results with previous studies on thermoelectric energy conversion in time-reversal broken mesoscopic
systems. In all previous studies, the charge and heat currents flowing out of the $P$ terminal are tuned to vanish by adjusting 
the chemical potential and temperature at the $P$ terminal (often called as a probe-terminal in mesoscopic physics). 
Under such constraints, there are effectively only one heat current and one electric current in the system. 
Thermoelectric transport is then described by a $2\times2$ Onsager matrix~\cite{saito,Balachandran,Udo-MT,BrandnerPRE,UdoPRL,BENENTI20171}. In this limit, the matrices $\hat{\mathcal M}_{ee}$,
$\hat{\mathcal M}_{eQ}$ and $\hat{\mathcal M}_{Qe}$ become scalar quantities. From the definition in
Eq.~(\ref{eq:lambda}), one finds that for such a set-up
\begin{equation}
\lambda_3^2=\lambda_1\lambda_2.
~\label{eq:2T_lambda}
\end{equation}
The above constraint is the main
limitation of previous studies, which is overcome in this work. As a consequence, the maximum efficiency in
our set-up can exceed that in previous set-ups, as shown in Fig.~3(a). In the figure, the black dot represent
the limit (\ref{eq:2T_lambda}) considered in previous studies. It is seen that the maximum efficiency can be 
improved by going beyond such a limit when $\lambda_1<\lambda_2$. Because of the power-efficiency trade-off,
the higher efficiency is achieved at lower output power, as shown in Fig.~3(b). Figs.~3(c) and 3(d) show the
maximal efficiency and the output power at such an efficiency. It is seen that large efficiency and power can
be simultaneous obtained when $\lambda_1>\lambda_2$.

%%%%%%%%%%%%%%%%%%%%%%%%%%%%%%%%%%%%%%%%%%%%%%%%%%%%%%%%%%%%%%%%%%
\begin{figure}[htb]
\begin{center}
\centering\includegraphics[width=4.2cm]{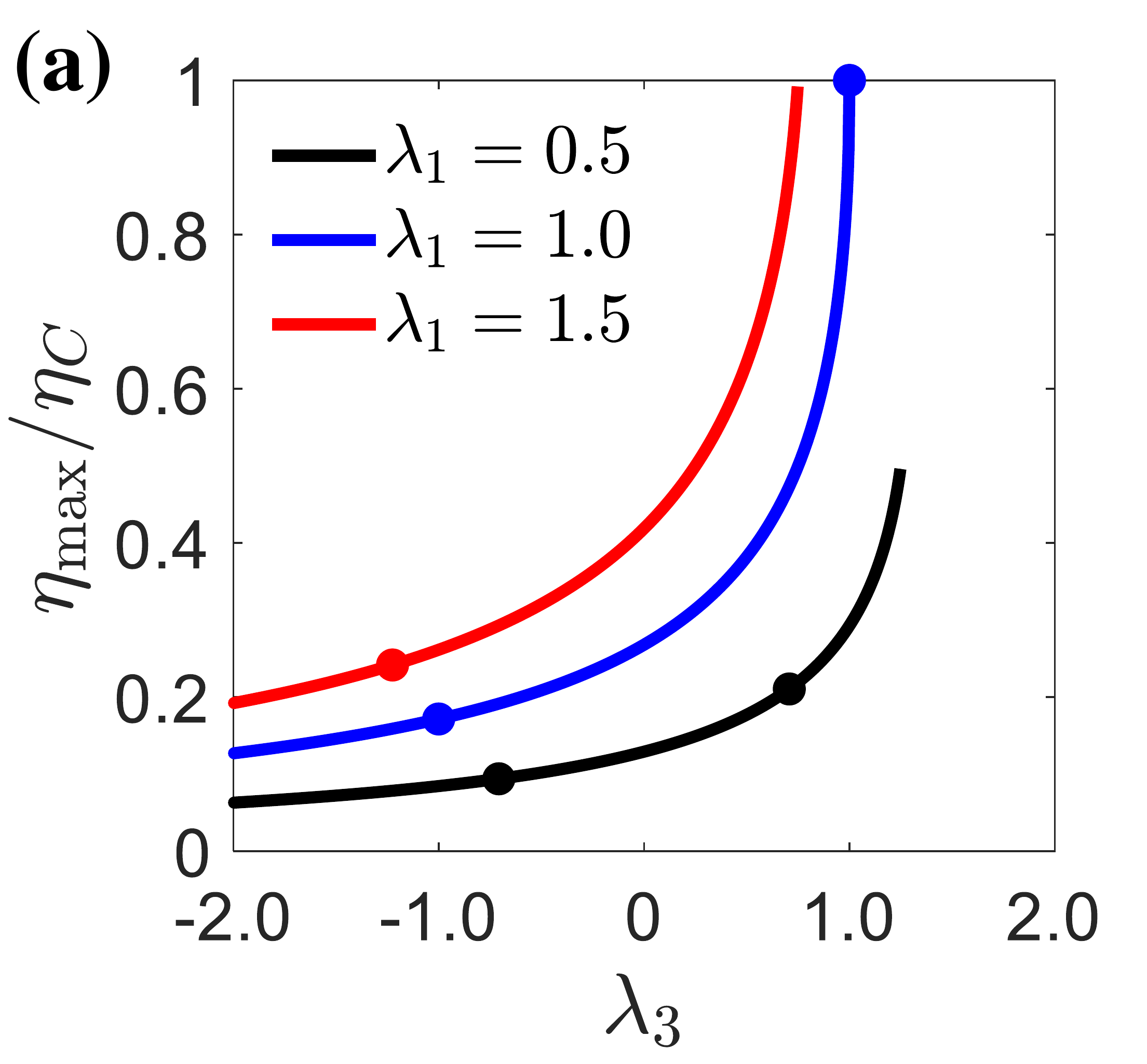}\hspace{0.2cm}\includegraphics[width=4.2cm]{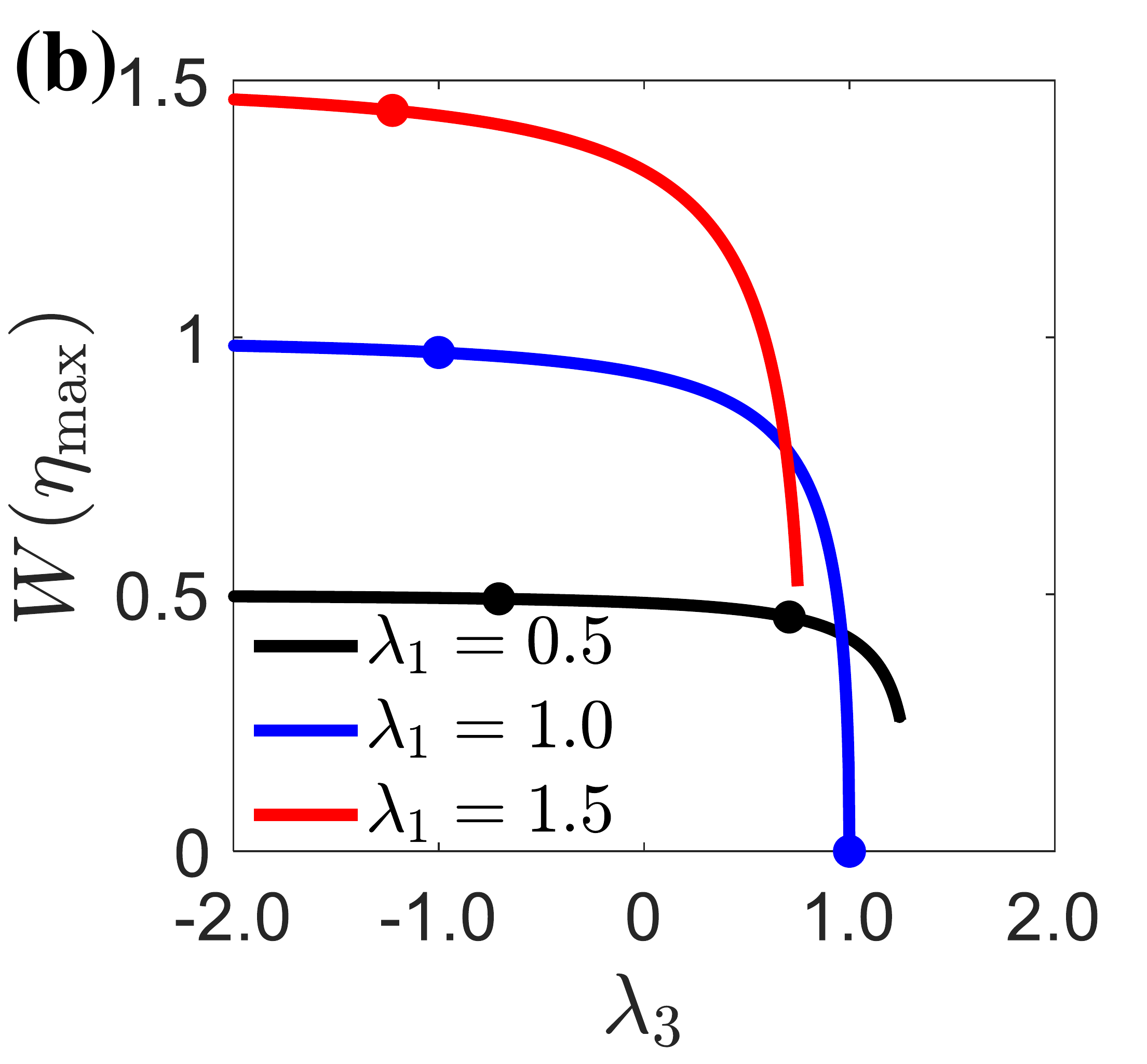}
\centering\includegraphics[width=4.2cm]{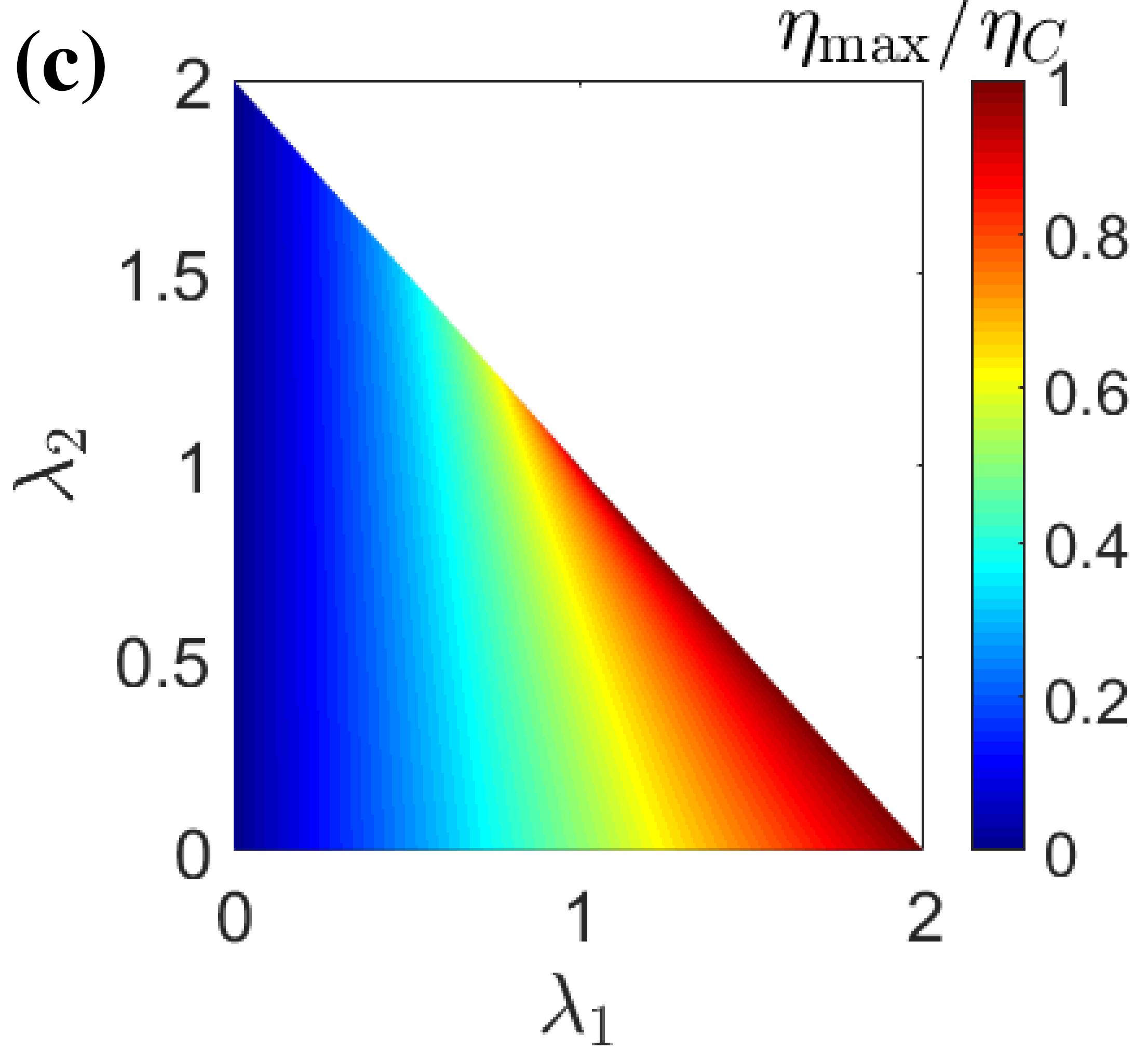}\hspace{0.2cm}\includegraphics[width=4.2cm]{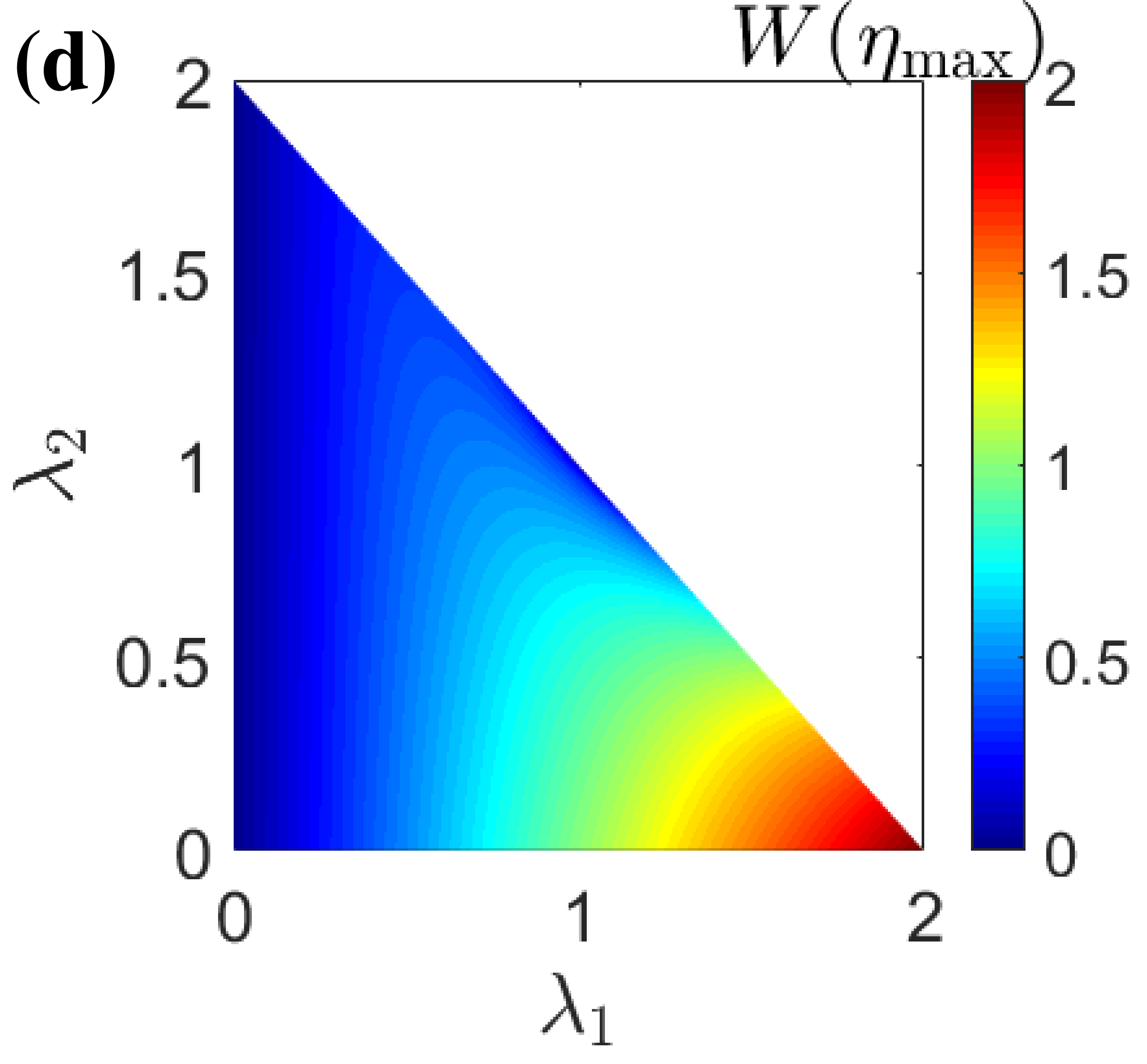}
\caption{(Color online) (a) $\eta_{\max}/\eta_C$ and (b) $W(\eta_{\max})$ as functions of $\lambda_3$ for different $\lambda_1$, where $\lambda_2=1$. The black dots represent the limit given by Eq.~\eqref{eq:2T_lambda}. (c) $\eta_{\max}$ and (d) $\eta(W_{\max})$ as functions of $\lambda_1$ and $\lambda_2$ for $\lambda_3=1$. The white region is forbidden by the second-law of thermodynamics. The unit of the output power is $W_0$. }
\label{fig:fig2}
\end{center}
\end{figure}
%%%%%%%%%%%%%%%%%%%%%%%%%%%%%%%%%%%%%%%%%%%%%%%%%%%%%%%%%%%%%%%%%%

\section{Upper bounds for efficiency and power}\label{upper-bound}
The bounds for the maximal efficiency $\eta_{max}$ and efficiency at the maximum power 
$\eta(W_{max})$ are reached at 
the reversible limit with $\lambda_1+\lambda_2+2\lambda_3=4$, leading to
\begin{equation}
\eta_{\max}|_{\rm{bound}}=\left\{
\begin{aligned}
& \eta_C\frac{\lambda_1}{\lambda_2}, \quad  \rm{if} \quad \lambda_1<\lambda_2, \\
& \eta_C, \quad\quad \,\rm{if} \quad   \lambda_1\ge\lambda_2.
\end{aligned}
\right.~\label{eq:bound_eta_m}
\end{equation}
The above results are presented graphically in Fig.~4(a) for various $\lambda_1$ and $\lambda_2$.
The upper bound for the efficiency at the maximum output power is
\begin{equation}
\eta(W_{\max})|_{\rm{bound}}=\eta_C\frac{\lambda_1}{\lambda_1+\lambda_2}.
~\label{eq:bound_eta_Wm}
\end{equation}
From the above, the Curzon-Ahlborn limit~\cite{Curzon-limit,Seifert2012} for the energy efficiency at maximum 
power $\eta=\eta_C/2$ can in principle be overcome for $\lambda_1>\lambda_2$. A particularly interesting regime is
when $\lambda_1\gg \lambda_2$ where both the maximal efficiency and the efficiency at maximum power 
are bounded by the Carnot efficiency.

Combining Eq.~\eqref{eq:bound_eta_m} and Eq.~\eqref{eq:bound_eta_Wm}, we find the ratio between those
bounds for energy efficiency, 
\begin{equation}
\frac{\eta_{\max}}{\eta(W_{\max})}|_{\rm{bound}}=\left\{
\begin{aligned}
& 1+\frac{\lambda_1}{\lambda_2}, \quad  \rm{if} \quad \lambda_1<\lambda_2, \\
& 1+\frac{\lambda_2}{\lambda_1}, \quad  \rm{if} \quad \lambda_1\ge\lambda_2.
\end{aligned}
\right.\label{eq:ratio_eta}
\end{equation}
Meanwhile, the ratio between those bounds for output power is given by
\begin{equation}
\frac{W(\eta_{\max})}{W_{\max}}|_{\rm{bound}}=\left\{
\begin{aligned}
& 1-\frac{\lambda_1}{\lambda_2}, \quad  \rm{if} \quad \lambda_1<\lambda_2, \\
& 1-\frac{\lambda_2}{\lambda_1}, \quad  \rm{if} \quad \lambda_1\ge\lambda_2.
\end{aligned}
\right.\label{eq:ratio_W}
\end{equation}
As presented graphically in Figs.~4(c) and 4(d) for various $\lambda_1$ and $\lambda_2$, the trade-off between the
optimal efficiency and power is significantly reduced when $\lambda_1\gg \lambda_2$, which implies that in this regime,
large energy efficiency and power can be obtained simultaneously.

%%%%%%%%%%%%%%%%%%%%%%%%%%%%%%%%%%%%%%%%%%%%%%%%%%%%%%%%%%%%%%%%%%
\begin{figure}[htb]
\begin{center}
\centering\includegraphics[width=4.2cm]{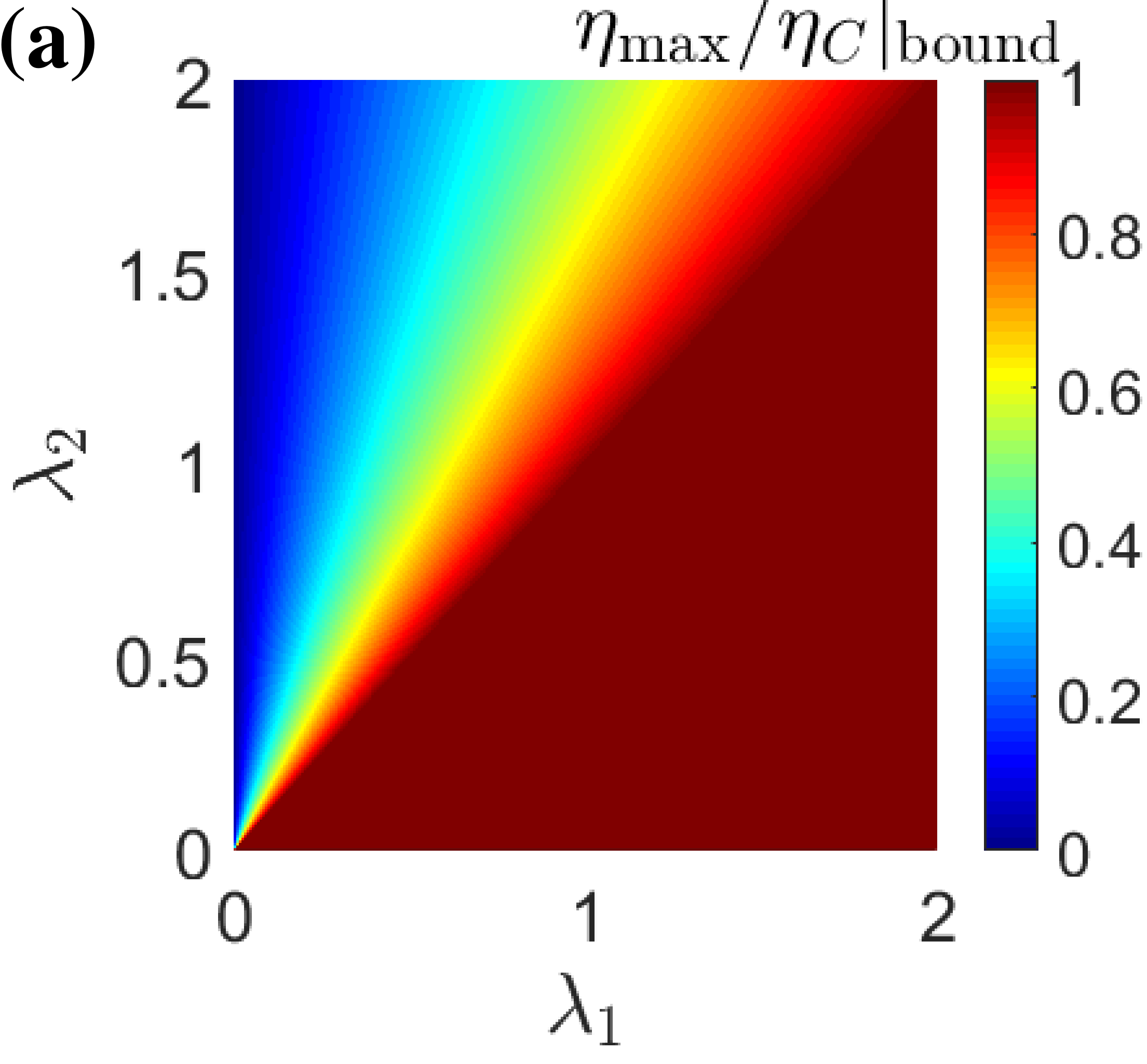}\hspace{0.2cm}\includegraphics[width=4.2cm]{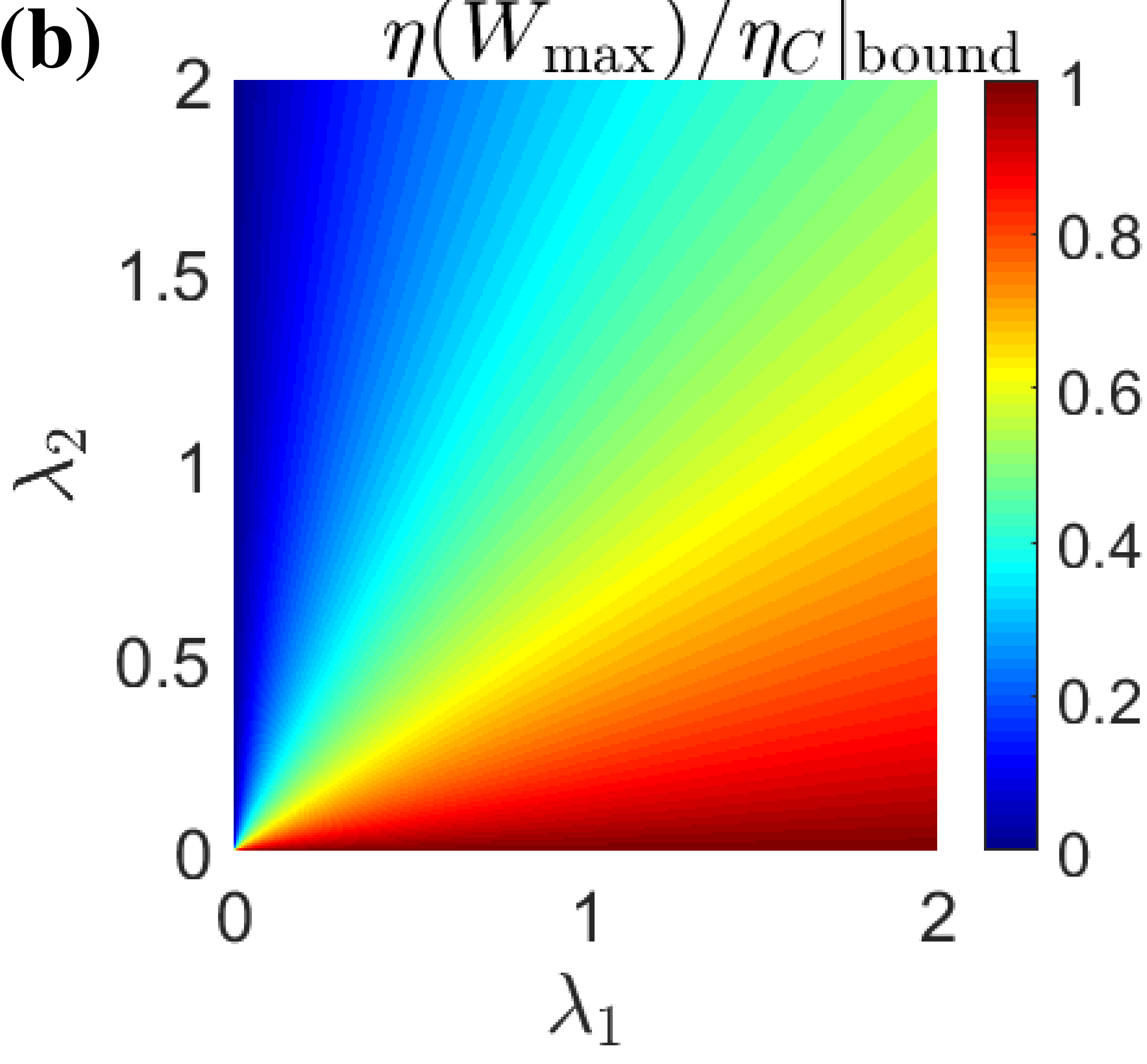}
\centering\includegraphics[width=4.2cm]{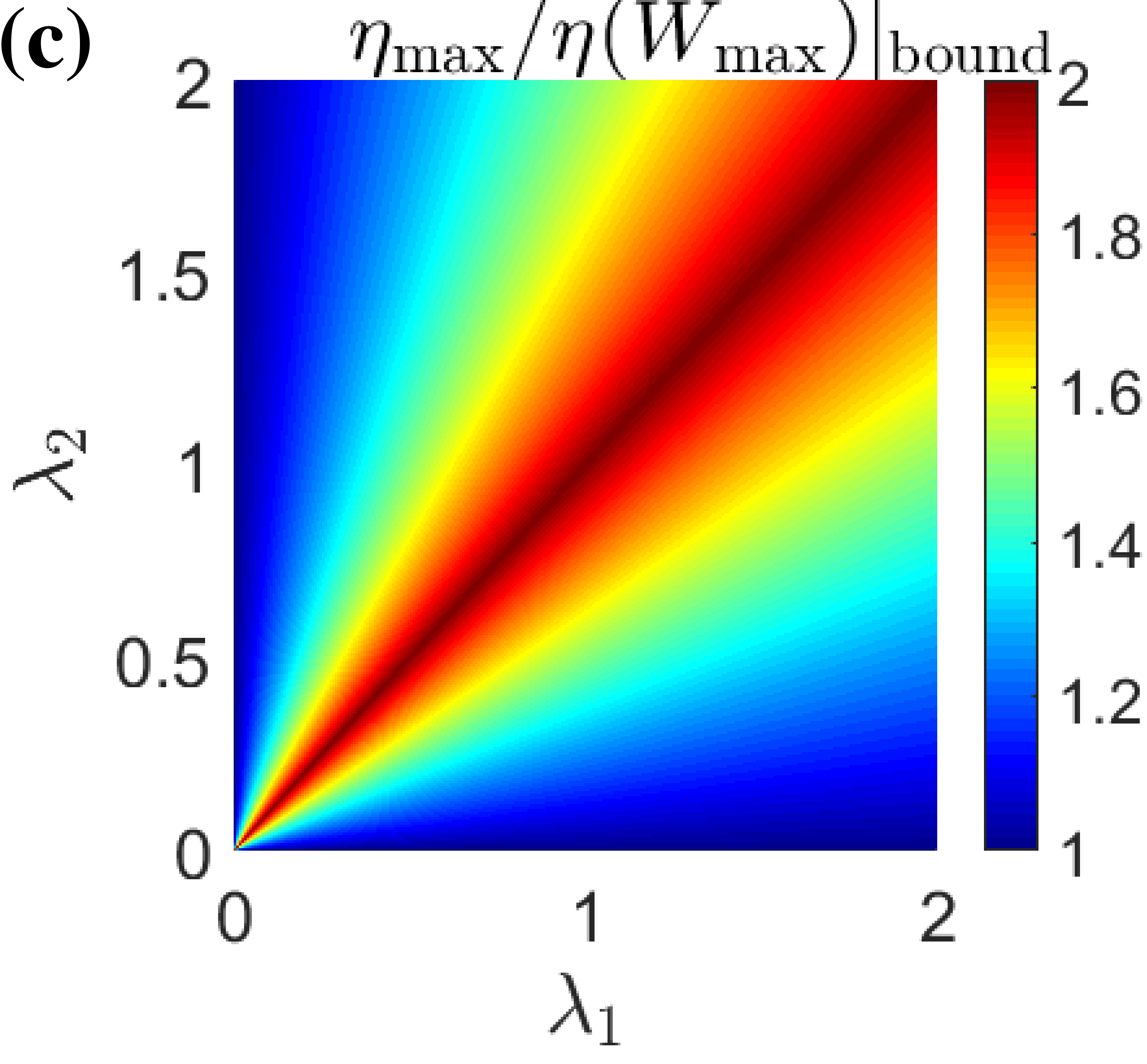}\hspace{0.2cm}\includegraphics[width=4.2cm]{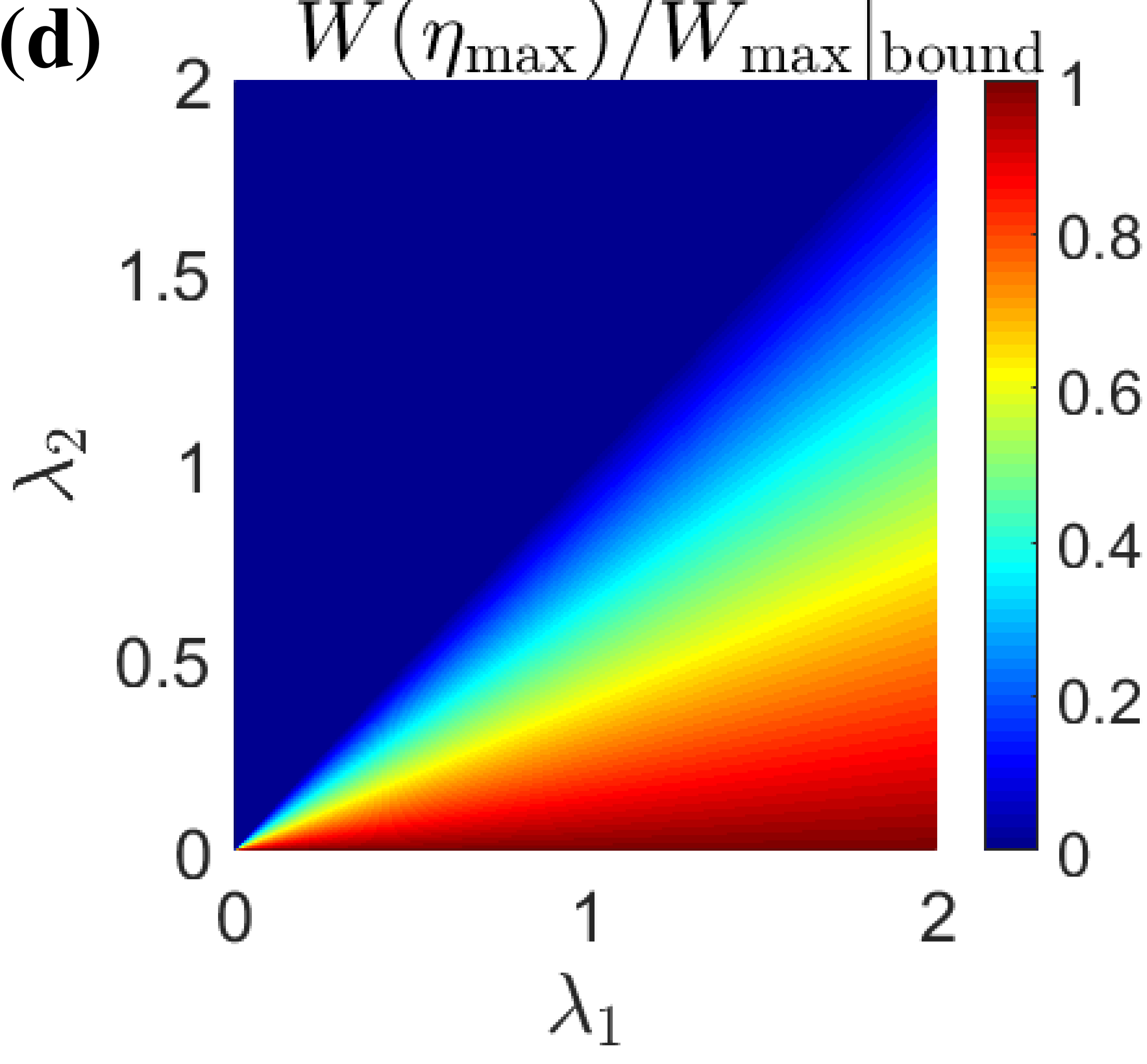}
\caption{(Color online) Bounds on efficiency and power. (a) $\eta_{\max}|_{\rm{bound}}$, (b) $\eta(W_{\max})|_{\rm{bound}}$, (c) $\eta_{\max}/\eta(W_{\max})|_{\rm{bound}}$ and (d) $W(\eta_{\max})/W_{\max}|_{\rm{bound}}$ as functions of $\lambda_1$ and $\lambda_2$.}
\label{fig:lambda}
\end{center}
\end{figure}
%%%%%%%%%%%%%%%%%%%%%%%%%%%%%%%%%%%%%%%%%%%%%%%%%%%%%%%%%%%%%%%%%%

\section{Linear-response coefficients in a noninteracting QD system} \label{non-system}

We now investigate the optimal efficiency and power with a concrete model. The model adopted here is the
three QDs model illustrated in Fig.~1, which has been studied extensively for the situations with only
one electric and one heat currents. By releasing such a constraint, the charge and heat transport are described by
the following equation,
\begin{equation}
\begin{pmatrix}
I^L_e \\
I^P_e \\
I_Q \\
\end{pmatrix}
=
\begin{pmatrix}
M_{11} & M_{12} & M_{13}  \\
M_{21} & M_{22} & M_{23}  \\
M_{31} & M_{32} & M_{33}  \\
\end{pmatrix}
\begin{pmatrix}
F^L_e \\
F^P_e \\
F_Q \\
\end{pmatrix},
\end{equation}
The coherent flow of charge and heat through a noninteracting QD system can be described using
the Landauer-B\"{u}tiker theory. The charge and heat currents flowing out of the left reservoir are given 
by~\cite{Sivan,butcher1990}
\begin{subequations}
\begin{align}
I^L_e & = \frac{2e}{h}\int dE\sum_i[({\mathcal T}_{iL}(E)f_L(E)-{\mathcal T}_{Li}(E)f_R(E))], \label{trans-LB}
\\
I_Q & = \frac{2}{h}\int dE\sum_i(E-\mu_L)[{\mathcal T}_{iL}(E)f_L(E)-{\mathcal T}_{Li}(E)f_R(E)],
\end{align}
\end{subequations}
where $f_i(E)=\{\exp[(E-\mu_i)/k_BT_i]+1\}^{-1}$ is the Fermi function and ${\mathcal T}_{ij}$ is the transmission probability from terminal $j$ to terminal $i$, $h$ is the Planck constant. The factor of two comes from the spin degeneracy of electrons. Analogous expression can be written for $I^P_e$, provided the label $L$ is substituted by $P$ in (\ref{trans-LB}).

The Onsager coefficients $M_{ij}$ are obtained from the linear expansion of the electronic currents $I_e^i$ ($i=L,P$)
and the heat current $I_Q$ in terms of the thermodynamic forces~\cite{Sivan,butcher1990},
\begin{equation}
\begin{aligned}
M_{11}&=\frac{2e^2}{hk_BT}\int_{-\infty}^\infty dE \sum_{i\ne L} {\mathcal T}_{Li}(E) F(E),\\
M_{12}&=-\frac{2e^2}{hk_BT}\int_{-\infty}^\infty dE {\mathcal T}_{LP}(E) F(E),\\
M_{13}&=M_{31}= \frac{2e}{hk_BT}\int_{-\infty}^\infty dE (E-\mu) \sum_{i\ne L} {\mathcal T}_{Li}(E) F(E),\\
M_{21}&=-\frac{2e^2}{hk_BT}\int_{-\infty}^\infty dE {\mathcal T}_{PL}(E) F(E),\\
M_{22}&=\frac{2e^2}{hk_BT}\int_{-\infty}^\infty dE \sum_{i\ne P} {\mathcal T}_{Pi}(E)F(E),\\
M_{23}&=-\frac{2e}{hk_BT}\int_{-\infty}^\infty dE(E-\mu){\mathcal T}_{PL}(E) F(E),\\
M_{32}&=-\frac{2e}{hk_BT}\int_{-\infty}^\infty dE(E-\mu){\mathcal T}_{LP}(E) F(E),\\
M_{33}&=\frac{2}{hk_BT}\int_{-\infty}^\infty dE (E-\mu)^2\sum_{i\ne L} {\mathcal T}_{Li}(E) F(E).\label{eq:Mij}
\end{aligned}
\end{equation}
where $F(E)\equiv\{4\cosh^2[(E-\mu)/k_BT]\}^{-1}$.

\begin{figure}[htb]
\begin{center}
\centering \includegraphics[width=4.2cm]{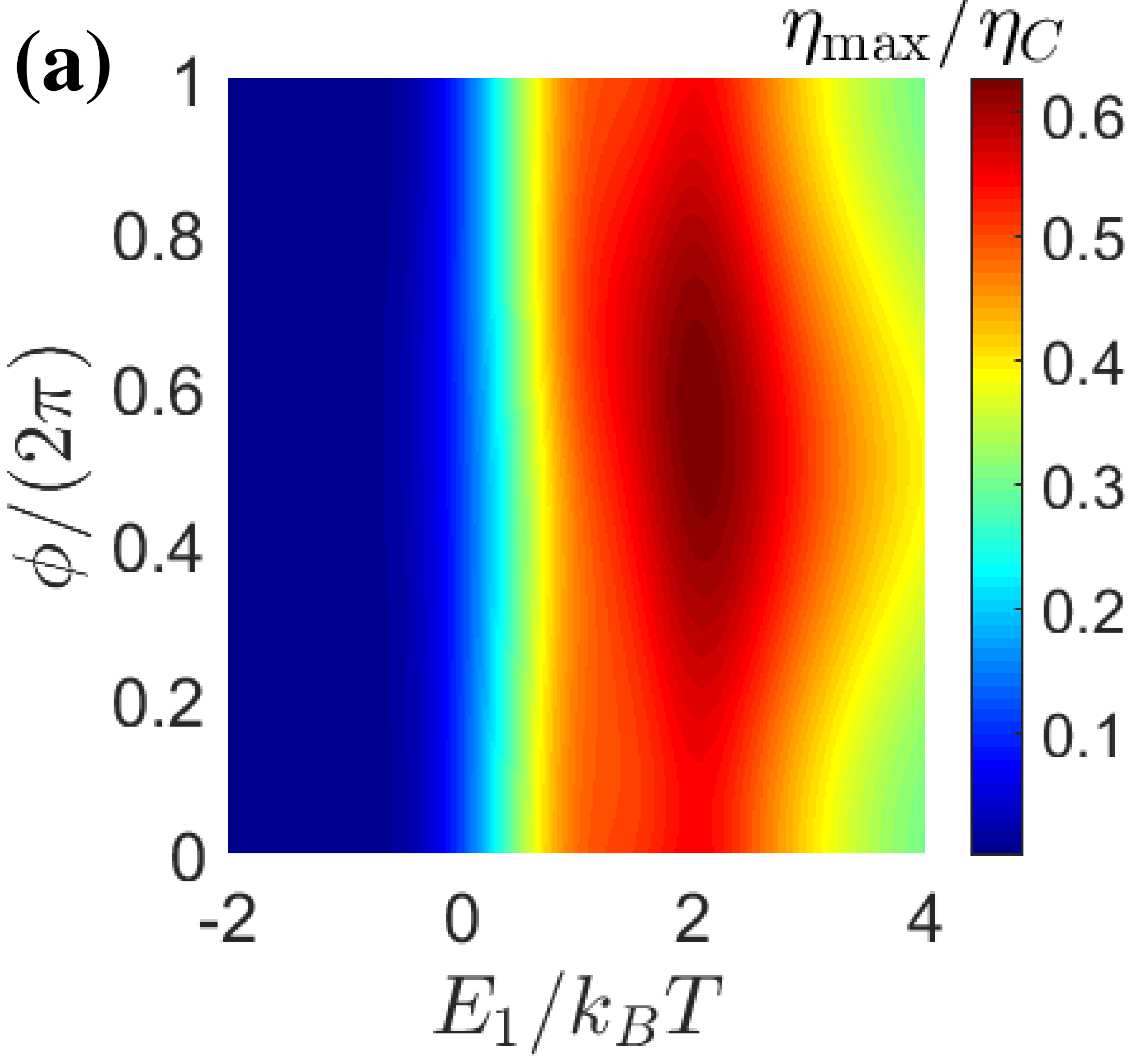}\hspace{0.2cm}\includegraphics[width=4.2cm]{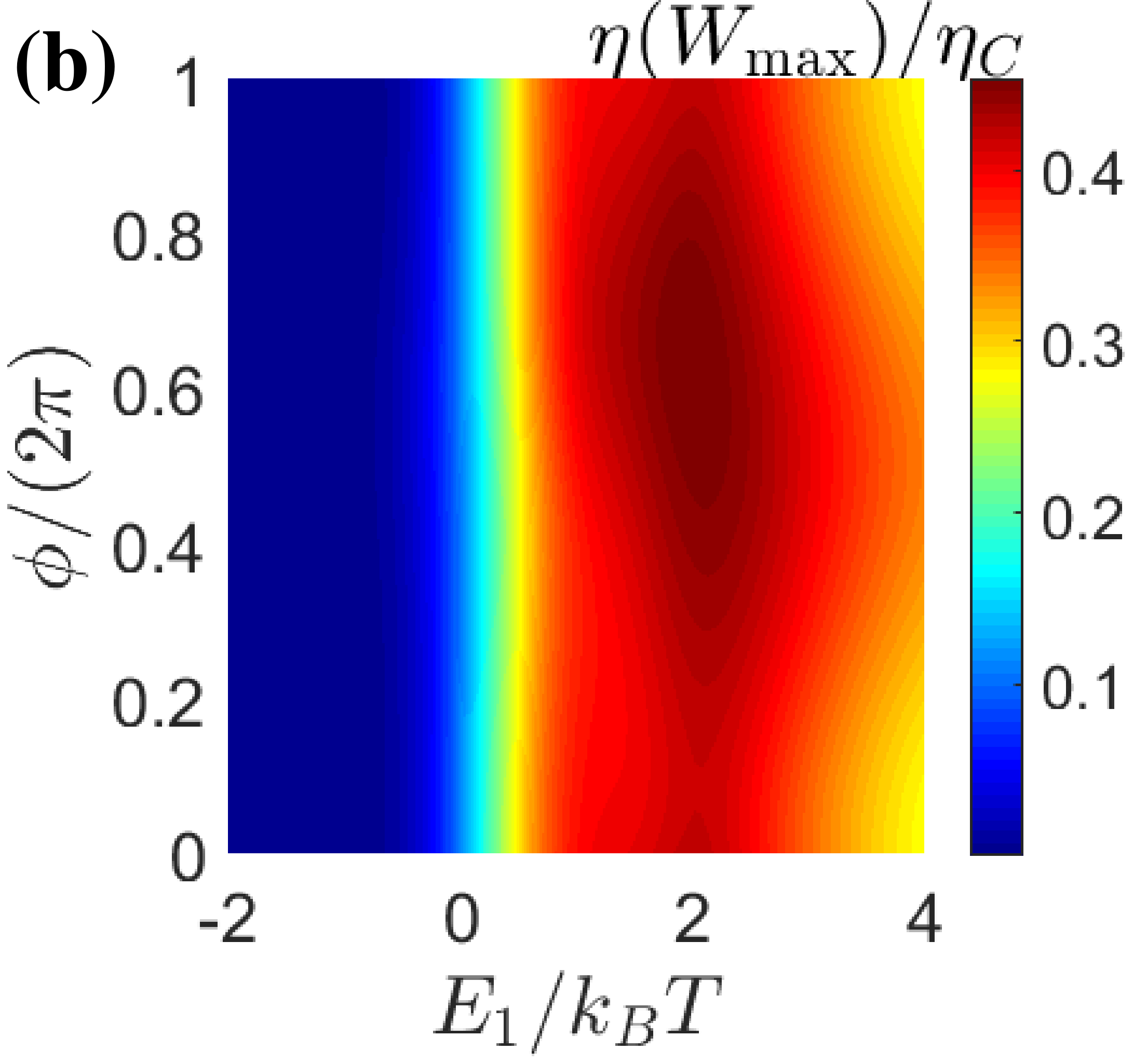}
\centering \includegraphics[width=4.2cm]{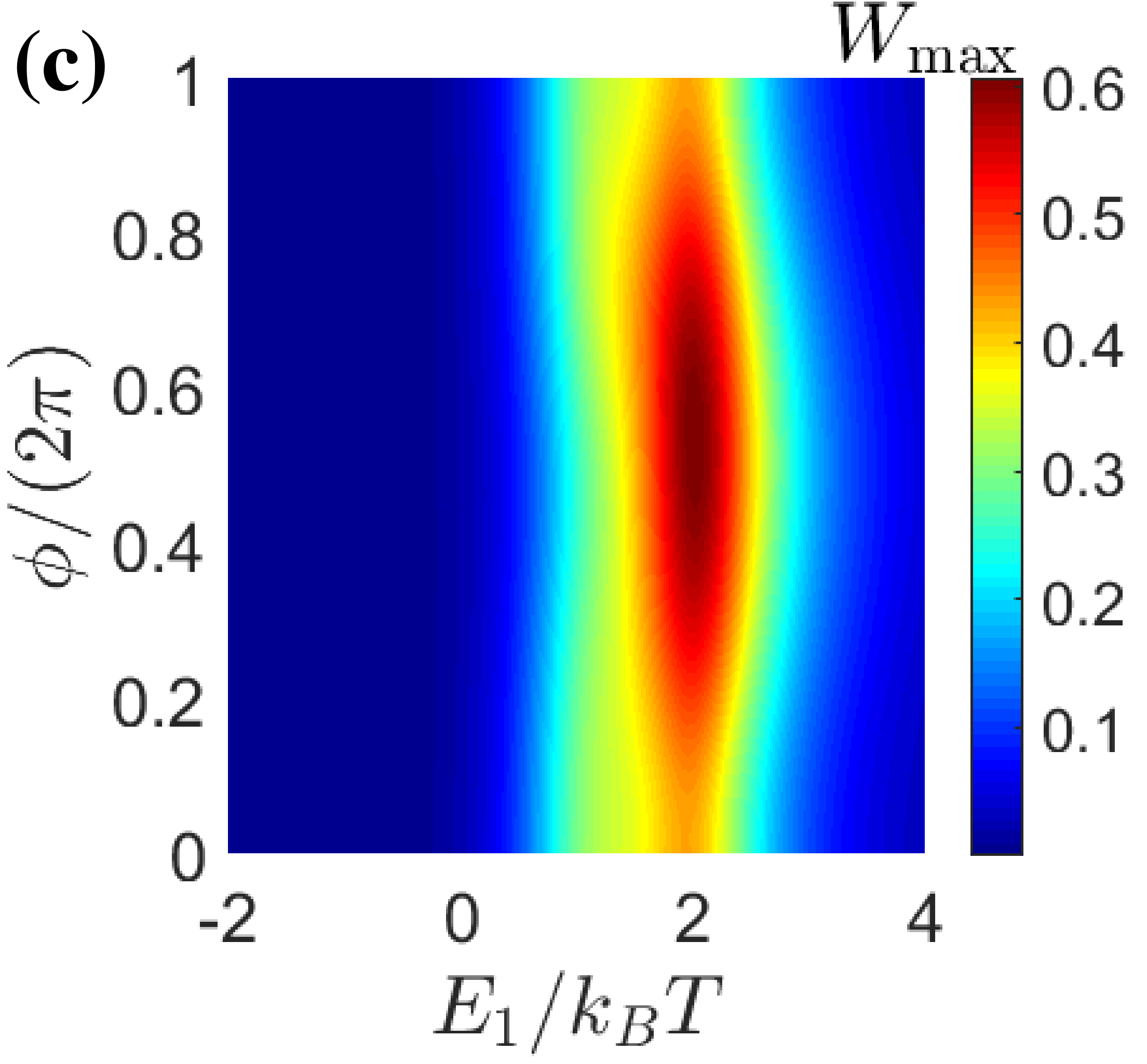}\hspace{0.2cm}\includegraphics[width=4.2cm]{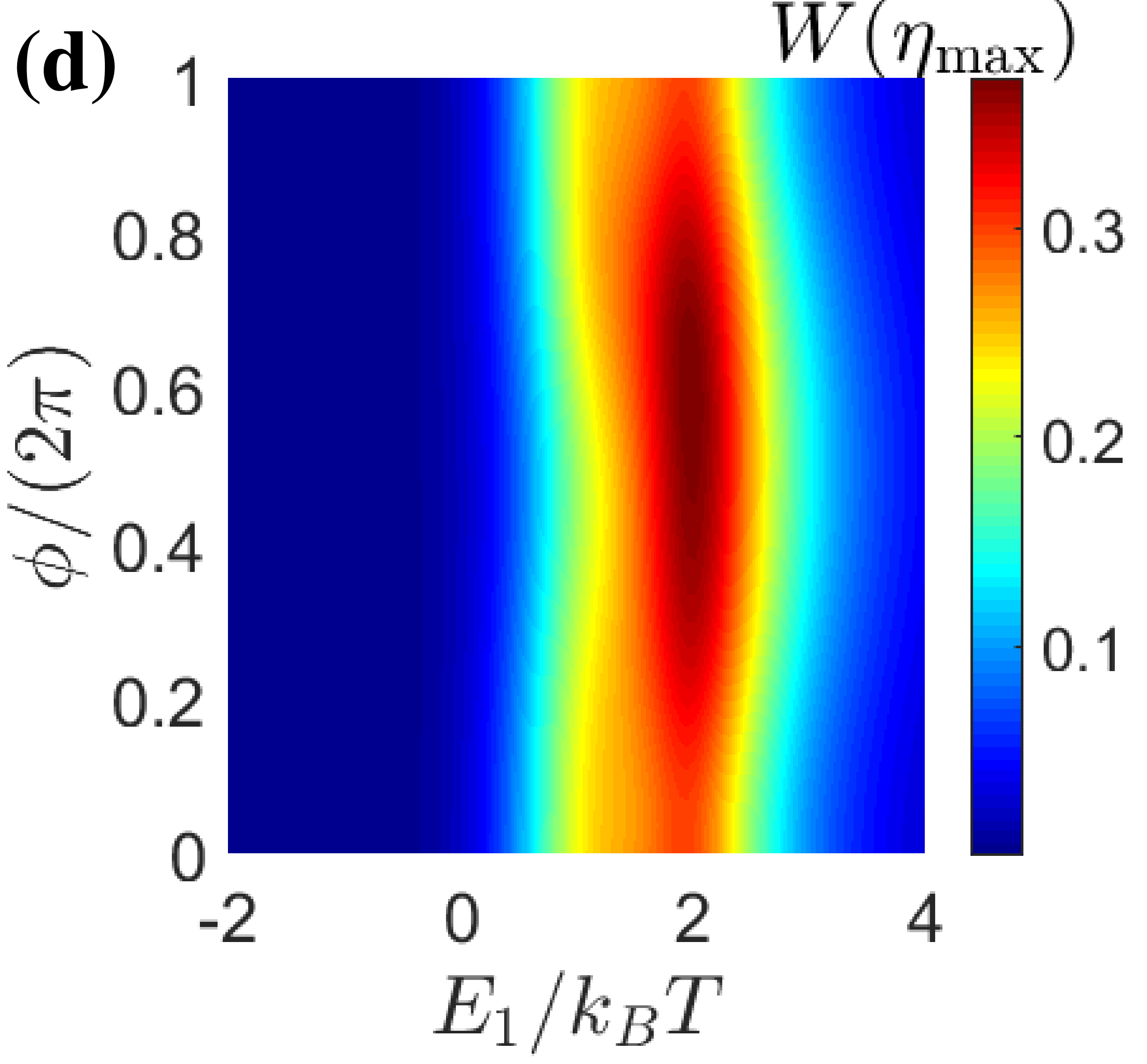}
\caption{(Color online) Optimal efficiency and power in a triple-QD thermoelectric engine. (a) $\eta_{\max}$, (b) $\eta(W_{\max})$, (c) $W_{\max}$, and (d) $W(\eta_{\max})$ as the functions of $E_1$ and $\phi$. The other parameters are $t=-0.2k_BT$, $\Gamma=0.5k_BT$, $\mu=0$, $E_2=1.0k_BT$ and $E_3=2.0k_BT$. }\label{fig:E1phi}
\end{center}
\end{figure}

The transmission probability ${\mathcal T}_{ij}(E)$ is calculated as ~\cite{Balachandran}
\begin{equation}
{\mathcal T}_{ij}={\rm Tr}[\Gamma_i(E)G(E)\Gamma_j(E)G^\dagger(E)],
\label{eq:Tpq}
\end{equation}
where the (retarded) Green's function for the QD system is $G(E)\equiv (E-H_{qd}-i\Gamma/2)^{-1}$, the damping rate $\Gamma=2\pi\sum_k|V_{ik}|^2\delta(\omega-\varepsilon_{ik})$ is assumed to be a constant for all three leads.

When an external magnetic field $\Phi$ is applied to the system, the laws of physics remain unchanged if time $t$ is replaced by $-t$, provided that simultaneously the magnetic field ${\Phi}$ is replaced by $-{\Phi}$. In this case, the transport coefficients meet the Onsager-Casimir relations~\cite{datta}
\begin{equation}
M_{ij}({\phi})=M_{ji}(-{\phi}).
\end{equation}
It is seen from Fig.~\ref{fig:E1phi} that the optimal efficiency and power vary strongly with the QD energy $E_1$ and the 
magnetic flux $\phi$. For these cases, the dependence on the QD energy is stronger than that on the magnetic flux.
The efficiency and power are large when $E_1\approx2k_BT$. The maximum efficiency $\eta_{\max}$ can reach $0.6\eta_C$. 
The results here reveal that a small external magnetic field can improve both the power and efficiency, when starting from
the time-reversal limit of $\phi=\pi$.

%%%%%%%%%%%%%%%%%%%%%%%%%%%%%%%%%%%%%%%%%%%%%%%%%%%%%%%%%%%%%%%%%%
\begin{figure}[htb]
\begin{center}
\centering \includegraphics[width=4.2cm]{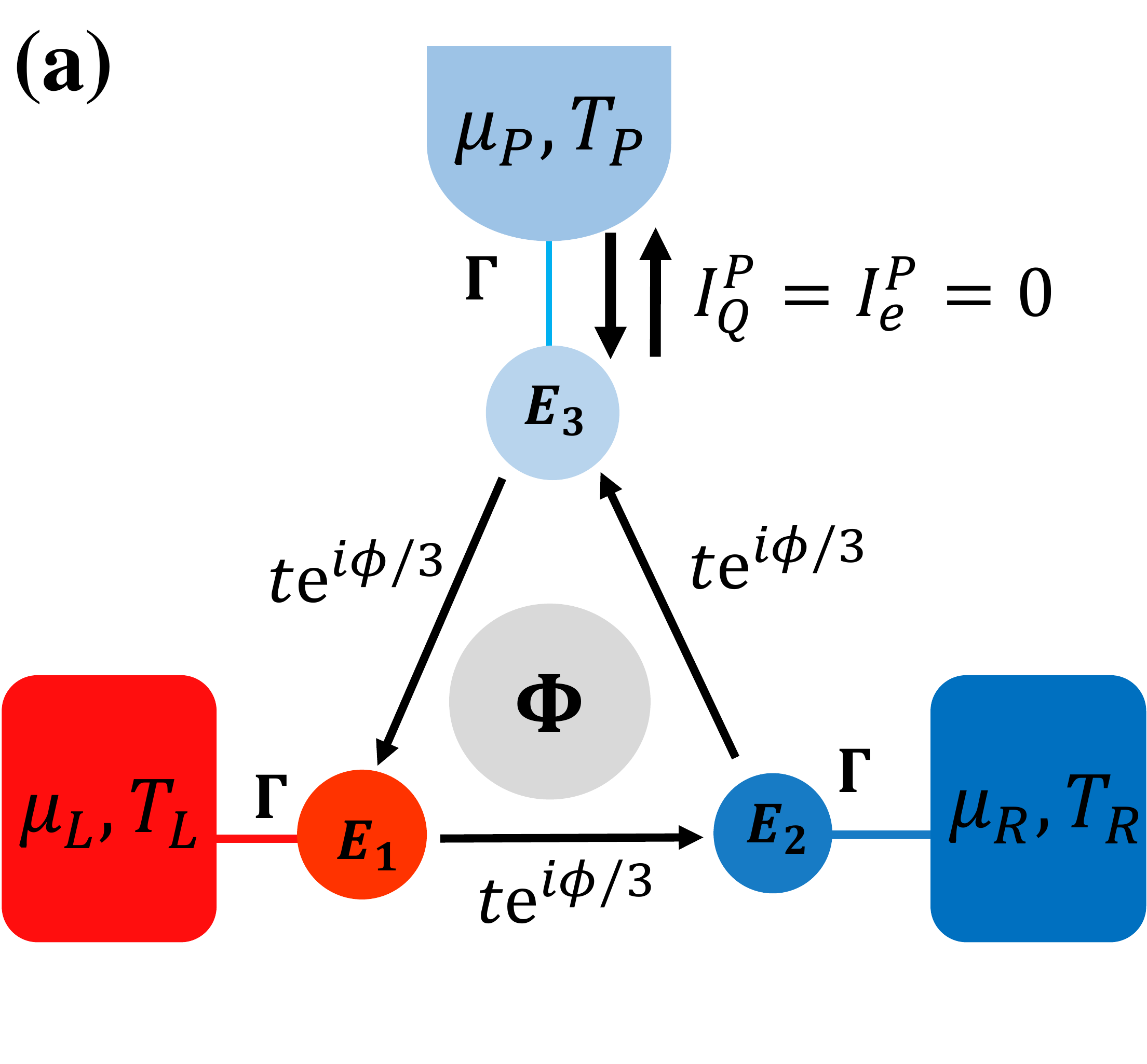}\hspace{0.2cm}\includegraphics[width=4.2cm]{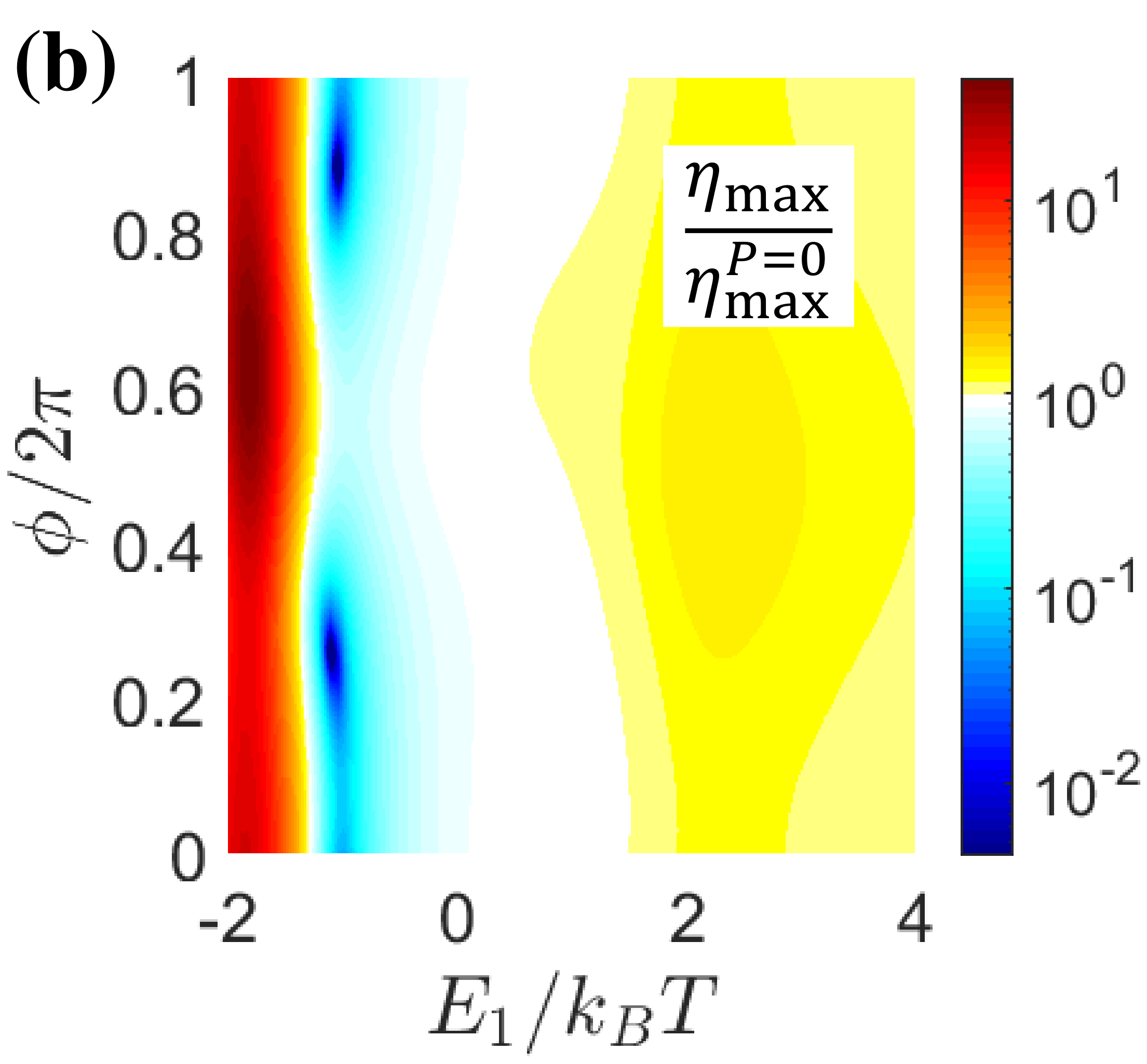}
\centering \includegraphics[width=4.2cm]{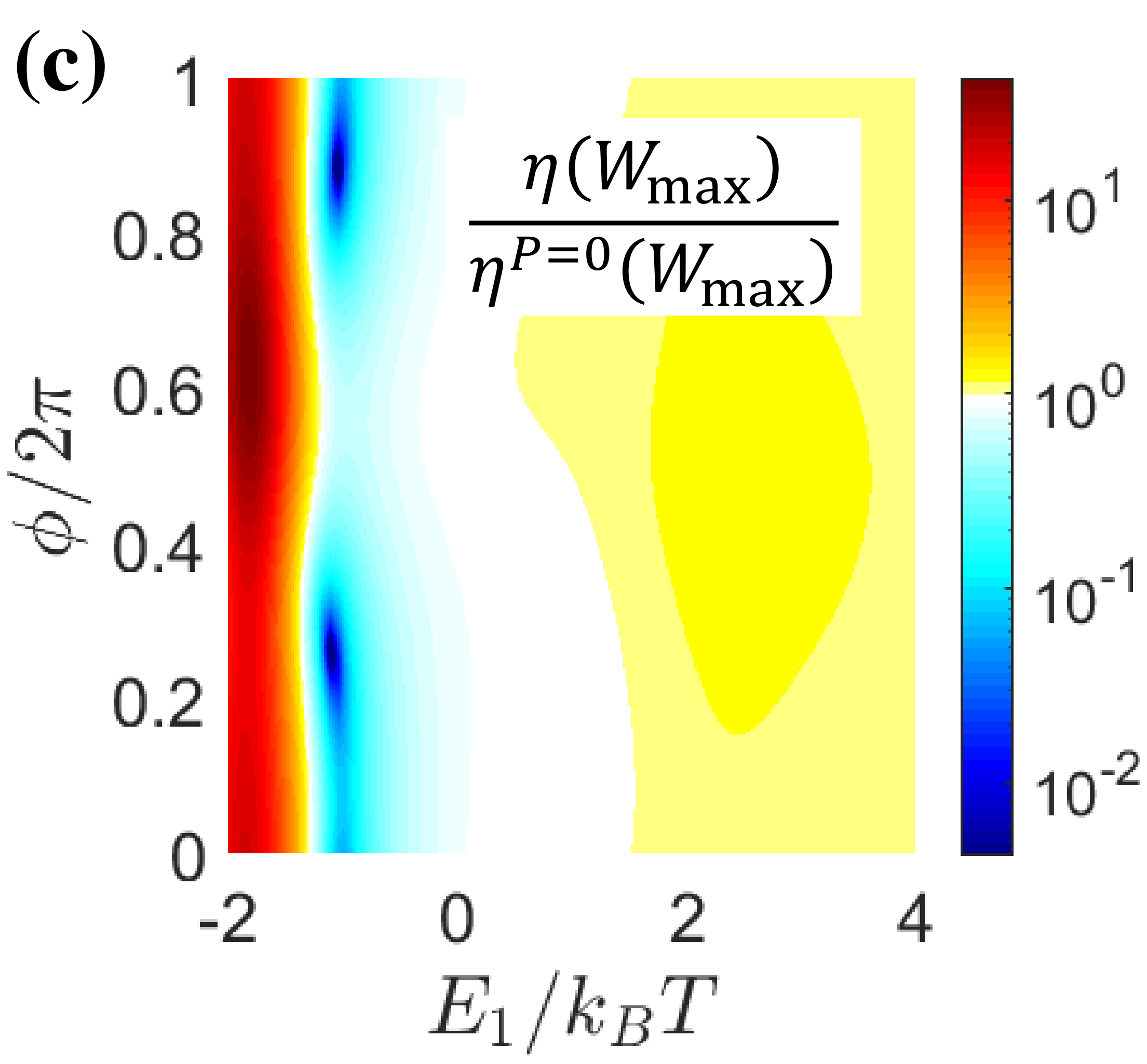}\hspace{0.2cm}\includegraphics[width=4.2cm]{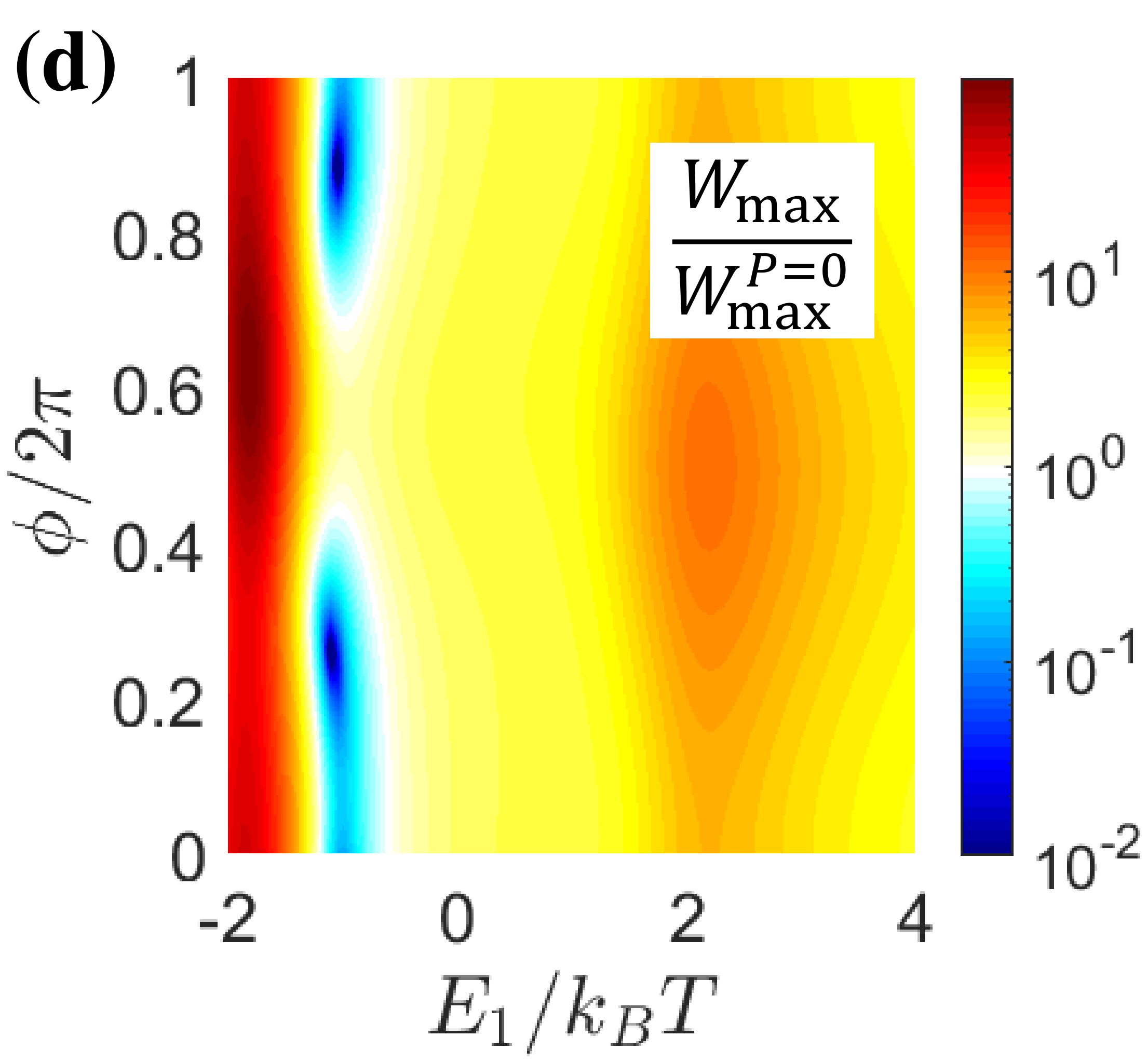}
\caption{(Color online) (a) Schematic of the triple-QD thermoelectric engine when the electric current $I_e^P$ and 
heat current $I_Q^P$ vanish (denoted as $P=0$ briefly). (b)-(d) Comparing the optimal efficiency and power for the $P=0$ limit
and the case with two output electric currents for various QD energy $E_1$ and magnetic flux $\phi$: (b) the maximal efficiency, (c) the efficiency at maximum power, and (d) the maximum output power. The other parameters are $t=-0.2k_BT$, $\mu=0$, $E_2=1.0k_BT$ and $E_3=2.0k_BT$.}\label{fig:2T-3T}
\end{center}
\end{figure}
%%%%%%%%%%%%%%%%%%%%%%%%%%%%%%%%%%%%%%%%%%%%%%%%%%%%%%%%%%%%%%%%%%

In Fig.~\ref{fig:2T-3T}, we compare explicitly the performance of our three-terminal quantum heat engine 
with the previously studied limit. The latter is illustrated in Fig.~\ref{fig:2T-3T}(a), where the heat and electric currents
flowing out of the $P$ terminal vanish by adjusting the chemical potential $\mu_P$ and temperature $T_P$. In this limit
(denoted as $P=0$ briefly)
there are only one electric and one heat currents, yielding the relation in Eq.~(\ref{eq:2T_lambda}).
As shown in Figs.~\ref{fig:2T-3T}(b), Fig.~\ref{fig:2T-3T}(c) and Fig.~\ref{fig:2T-3T}(d), the maximal efficiency, the efficiency 
at maximum power, and the maximum output power can be significantly improved by releasing the limit of $P=0$. Our 
quantum heat engine with two output electric currents demonstrate superior efficiency and power for a large range of
parameters.

\section{Conclusion}\label{conclusion}
In conclusion, we derived the optimal efficiency and power, and their trade-off relations for a three-terminal thermoelectric
engine with two output electric currents. These results go beyond previous studies with time-reversal symmetry~\cite{JiangPRE}, 
and the time-reversal broken systems~\cite{Saito2011,StrongBounds} with only one electric current, revealing universalities 
in multi-terminal thermoelectric energy conversion differing from the existing theories. Numerical calculations for a 
triple-QD thermoelectric engine show that the efficiency and power can be substantially improved for the set-ups 
with two output electric currents compared with previous set-ups with only one electric current. We also find regimes 
where the energy efficiency and output power can be optimized at close conditions. Our results offer 
useful guidelines for the search of high-performance thermoelectric systems in the mesoscopic regime, with particular
emphasizes on multi-terminal set-ups with multiple output electric currents.

\section*{Acknowledgment}
J.L., Y.L., R.W., and J.-H.J. acknowledge support from the National Natural Science Foundation of China (NSFC Grant No. 11675116) and a Project Funded by the Priority Academic Program Development of Jiangsu Higher Education Institutions (PAPD). C.W. is supported by the National Natural Science Foundation of China under Grant No. 11704093.

\bibliography{trade-off}

\end{document}